%% file: ms.tex
\documentclass{article}

\usepackage{arxiv}

\usepackage[utf8]{inputenc} 
\usepackage[T1]{fontenc}    
\usepackage{hyperref}       
\usepackage{url}            
\usepackage{booktabs}       
\usepackage{amsfonts}       
\usepackage{nicefrac}       
\usepackage{microtype}      
\usepackage{lipsum}		
\usepackage{graphicx}
\usepackage[numbers]{natbib}
\usepackage{amsmath}
\usepackage{mathrsfs}
\usepackage{multirow}

\title{The transformer earthquake alerting model: A new versatile approach to earthquake early warning}

\author{Jannes Münchmeyer$^{1, 2, *}$, Dino Bindi$^{1}$, Ulf Leser$^{2}$, Frederik Tilmann$^{1, 3}$\\
$^1$ Deutsches GeoForschungsZentrum GFZ, Potsdam, Germany\\
$^2$ Institut für Informatik, Humboldt-Universität zu Berlin, Berlin, Germany\\
$^3$ Insitut für geologische Wissenschaften, Freie Universität Berlin, Berlin, Germany\\
$^*$ To whom correspondence should be addressed: \url{munchmej@gfz-potsdam.de}
}

\renewcommand{\headeright}{}
\renewcommand{\undertitle}{}
\renewcommand{\shorttitle}{The transformer earthquake alerting model}

\begin{document}
\maketitle

\begin{abstract}
Earthquakes are major hazards to humans, buildings and infrastructure.
Early warning methods aim to provide advance notice of incoming strong shaking to enable preventive action and mitigate seismic risk.
Their usefulness depends on accuracy, the relation between true, missed and false alerts, and timeliness, the time between a warning and the arrival of strong shaking.
Current approaches suffer from apparent aleatoric uncertainties due to simplified modelling or short warning times.
Here we propose a novel early warning method, the deep-learning based transformer earthquake alerting model (TEAM), to mitigate these limitations.
TEAM analyzes raw, strong motion waveforms of an arbitrary number of stations at arbitrary locations in real-time, making it easily adaptable to changing seismic networks and warning targets.
We evaluate TEAM on two regions with high seismic hazard, Japan and Italy, that are complementary in their seismicity.
On both datasets TEAM outperforms existing early warning methods considerably, offering accurate and timely warnings.
Using domain adaptation, TEAM even provides reliable alerts for events larger than any in the training data, a property of highest importance as records from very large events are rare in many regions.
\end{abstract}


\section{Introduction}

The concept of earthquake early warning has been around for over a century, but the necessary instrumentation and methodologies have only been developed in the last three decades \citep{allenStatusEarthquakeEarly2009,allenEarthquakeEarlyWarning2019}.
Early warning systems aim to raise alerts if shaking levels likely to cause damage are going to occur.
Existing methods split into two main classes: source estimation based and propagation based.
The former, like EPIC \citep{chungOptimizingEarthquakeEarly2019} or FINDER \citep{boseFinDerImprovedRealtime2018}, estimate the source properties of an event, i.e., its location or fault extent and magnitude, and then use a ground motion prediction equation (GMPE) to infer shaking at target sites.
They provide long warning times, but incur a large apparent aleatoric uncertainty due to simplified assumptions in the source estimation and in the GMPE \citep{koderaPropagationLocalUndamped2018}.
Propagation based methods, like PLUM \citep{koderaPropagationLocalUndamped2018}, infer the shaking at a given location from measurements at nearby seismic stations.
Predictions are more accurate, but warning times are reduced, as warnings require measurements of strong shaking at nearby stations \citep{meierHowOftenCan2020}.

Recently, machine learning methods, particularly deep learning methods, have emerged as a tool for fast assessment of earthquakes.
Under certain circumstances, they led to improvements in various tasks, e.g., estimation of magnitude \citep{lomaxInvestigationRapidEarthquake2019,mousaviMachineLearningApproachEarthquake2020}, location  \citep{kriegerowskiDeepConvolutionalNeural2019,mousaviBayesianDeepLearningEstimationEarthquake2019} or peak ground acceleration (PGA) \citep{jozinovicRapidPredictionEarthquake2020}.
Nonetheless, no existing method is applicable to early warning because they lack real-time capabilities, instead requiring fixed waveform windows after the P arrival.
Furthermore, the existing methods are restricted in terms of their input stations, as they use either a single seismic station as input \citep{lomaxInvestigationRapidEarthquake2019,mousaviMachineLearningApproachEarthquake2020} or a fixed set of seismic stations, that needs to be defined at training time \citep{kriegerowskiDeepConvolutionalNeural2019,jozinovicRapidPredictionEarthquake2020,otakeDeepLearningModel2020}.
While single station approaches miss out on a considerable amount of information obtainable from combining waveforms from different sources, fixed stations approaches have limited adaptability to changing networks.
The latter is of particular concern as for large, dense networks the stations of interest, i.e., the stations closest to an event, will change on a per-event basis.
Finally, existing methods systematically underestimate the strongest shaking and the highest magnitudes, as these are rare and therefore underrepresented in the training data (Fig. 6, 8 in \citep{jozinovicRapidPredictionEarthquake2020}, Fig. 3, 4 in \citep{mousaviMachineLearningApproachEarthquake2020}).
However, early warning systems must also be able to provide reliable warnings for earthquakes larger than any previously seen in a region.

Here, we present the transformer earthquake alerting model (TEAM), a deep learning method for early warning, combining the advantages of both classical early warning strategies while avoiding the deficiencies of prior deep learning approaches.
We evaluate TEAM on two data sets from regions with high seismic hazard, namely Japan and Italy.
Due to their complementary seismicity, this allows to evaluate the capabilities of TEAM across scenarios.
We compare TEAM to two state-of-the-art warning methods, of which one is prototypical for source based warning and one for propagation based warning.

\section{Data and Methods}

\subsection{Data}

\begin{figure}
 \centering
 \includegraphics[width=\textwidth]{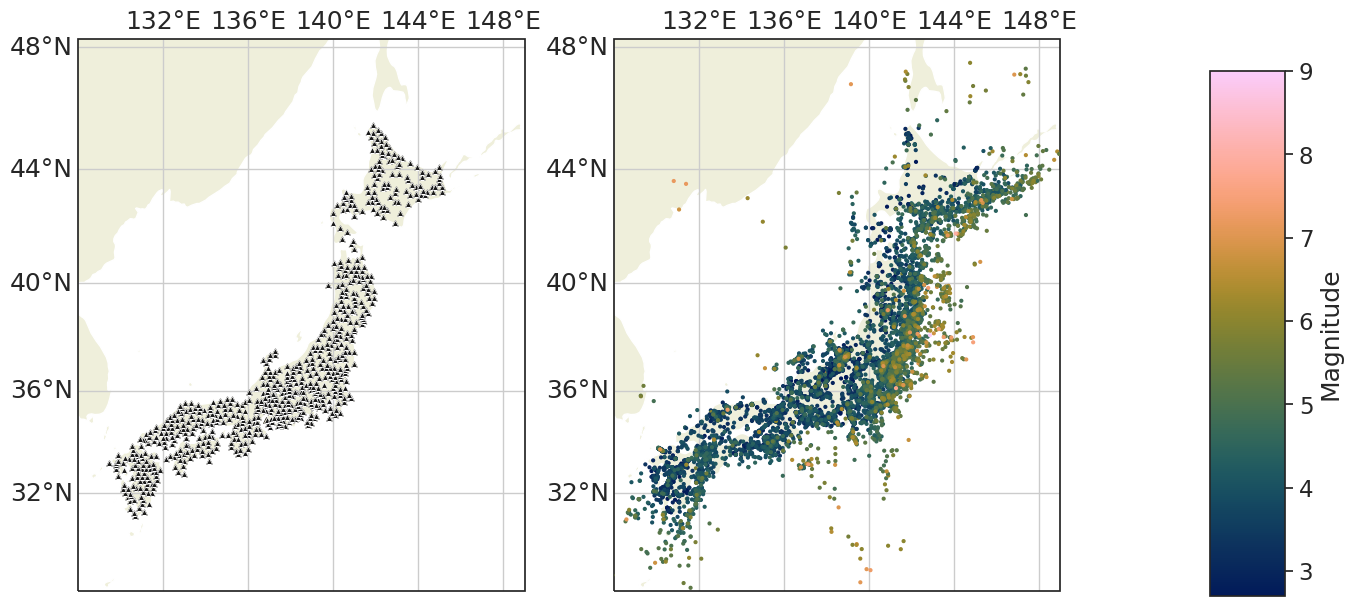}
 \caption{Map of the station (left) and event (right) distribution in the Japan dataset. Stations are shown as black triangles, events as dots. The event color encodes the event magnitude. There are $\sim$20~additional events far offshore, which are outside the displayed map region in the catalog.}
 \label{fig:map_japan}
\end{figure}

\begin{figure}
 \centering
 \includegraphics[width=\textwidth]{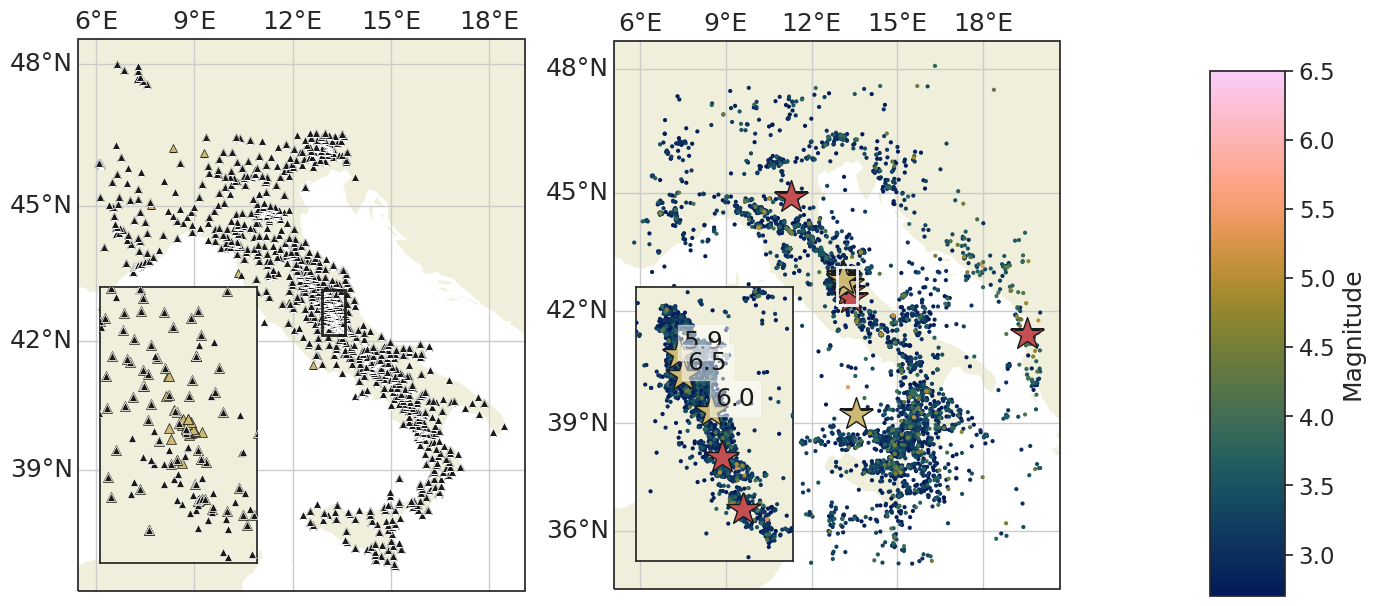}
 \caption{Map of the station (left) and event (right) distribution in the Italy dataset. Stations present in the training set are shown as black triangles, while stations only present in the test set are shown as yellow triangles. Events are shown as dots with the color encoding the event magnitude. All events with magnitudes above 5.5 are shown as stars. The red stars indicate large training events, while the yellow stars indicate large test events. The inset shows the central Italy region with intense seismicity and high station density in the test set. Moment magnitudes for the largest test events are given in the inset.}
 \label{fig:map_italy}
\end{figure}

For our study we use two nation scale datasets from highly seismically active regions with dense seismic networks, namely Japan (13,512 events, years 1997-2018, Figure~\ref{fig:map_japan}) and Italy (7,055 events, years 2008-2019, Figure~\ref{fig:map_italy}).
Their seismicity is complementary, with predominantly subduction plate interface or Wadati-Benioff zone events for Japan, many of them offshore, and shallow, crustal events for Italy.
We split both datasets into training, development and test sets with ratios of 60:10:30.
We employ an event-wise split, i.e., all records for a particular event will be assigned to the same subset.
We do not explicitly split station-wise but due to temporary deployments there are a few stations in the test set which have no records in the training set (Figure \ref{fig:map_italy}).
We use the training set for model training, the development set for model selection, and the test set only for the final evaluation.
We split the Japan dataset chronologically, yielding the events between August 2013 and December 2018 as test set.
For Italy, we test on all events in 2016, as these are of particular interest, encompassing most of the central Italy sequence with the $M_W$=6.2 and $M_w$=6.5 Norcia events \citep{dolce20162017Central2018}.
Especially the latter event is notably larger than any in the training set ($M_w = 6.1$ L'Aquila event in 2007), thereby challenging the extrapolation capabilities of TEAM.

Both datasets consist of strong motion waveforms.
For Japan each station comprises two sensors, one at the surface and one borehole sensor, while for Italy only surface recordings are available.
As the instrument response in the frequency band of interest is flat, we do not restitute the waveforms, but only apply a gain correction.
This has the advantage that it can trivially be done in real-time.
The data and preprocessing are further described in the supplement text S1.

\subsection{The transformer earthquake alerting model}

\begin{figure}
 \includegraphics[width=\textwidth]{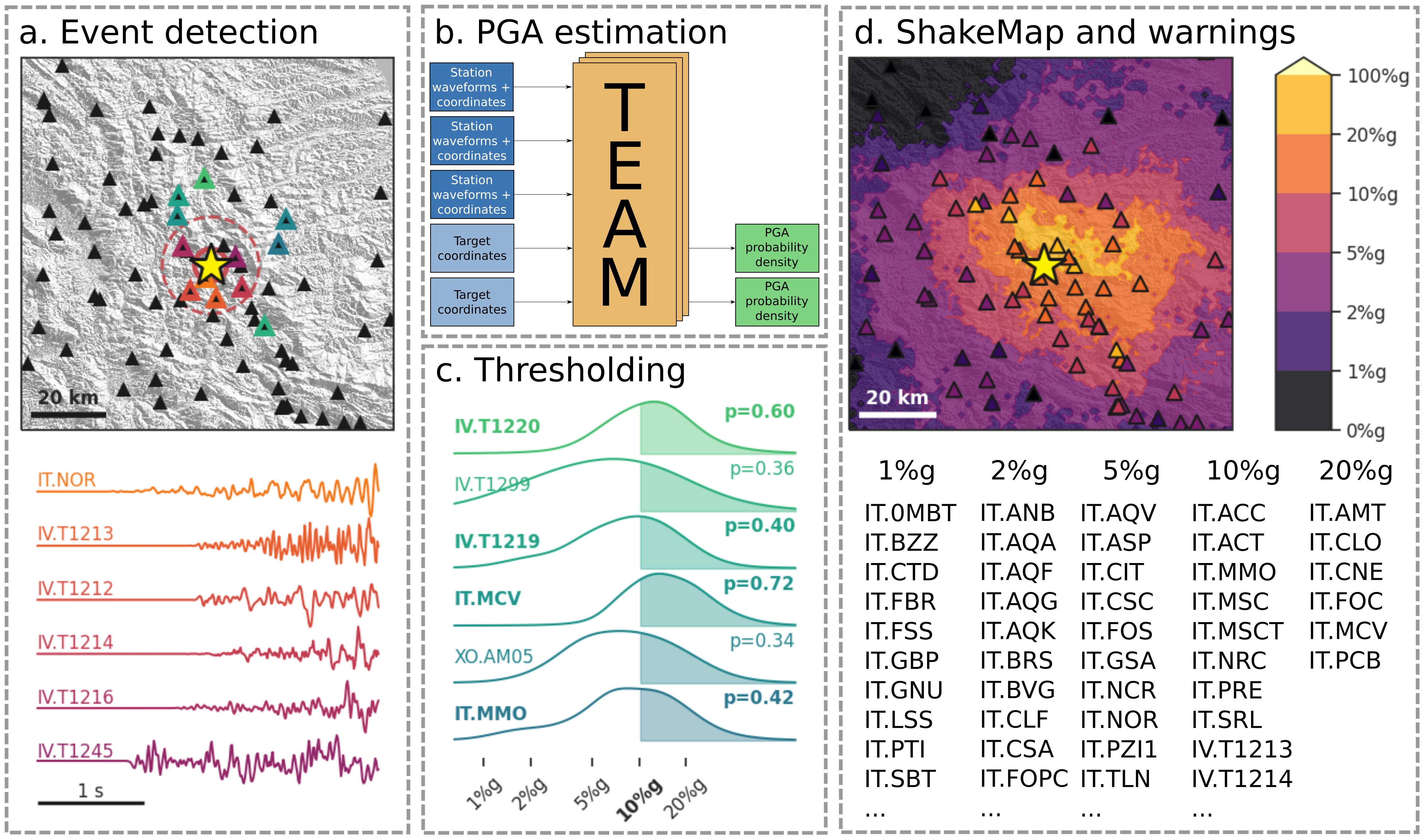}
 \caption{Schematic view of TEAM's early warning workflow for the October 2016 Norcia event ($M_w=6.5$) 2.5~s after the first P wave pick ($\sim$3.5~s after origin time).
 \textbf{a.} An event is detected through triggering at multiple seismic stations.
 The waveform colors correspond to the stations highlighted with orange to magenta outlines. 
 The circles indicate the approximate current position of P (dashed) and S (solid) wavefronts.
 \textbf{b.} TEAM's input are raw waveforms and station coordinates; it estimates probability densities for the PGA at a target set.
 A more detailed TEAM overview is given in Figure S1.
 \textbf{c.} The exceedance probabilities for a fixed set of PGA thresholds are calculated based on the estimated PGA probability densities.
 If the probability exceeds a threshold $\alpha$, a warning is issued.
 The figure visualizes a 10\%g PGA level with $\alpha=0.4$, resulting in warnings for the stations highlighted.
 The colors correspond to the stations with green outlines in a.
 \textbf{d.}
 The real-time shake map shows the highest PGA levels for which a warning is issued. 
 Stations are colored according to their current warning level.
 The table lists all stations for which warnings have already been issued.
 }
 \label{fig:scheme}
\end{figure}

The early warning workflow with TEAM encompasses three separate steps (Figure \ref{fig:scheme}): event detection, PGA estimation and thresholding.
We do not further consider the event detection task here, as it forms the base of all methods discussed and affects them similarly.
The PGA estimation, resulting in PGA probability densities for a given set of target locations, is the heart of TEAM and described in detail below.
In the last step, thresholding, TEAM issues warnings for each target locations where the predicted exceedance probability $p$ for fixed PGA thresholds surpasses a predefined probability $\alpha$.

TEAM conducts end-to-end PGA estimation: its input are raw waveforms, its output predicted PGA probability densities.
There are no intermediate representations in TEAM that warrant an immediate geophysical interpretation.
The PGA assessment can be subdivided into three components: feature extraction, feature combination, and density estimation (Figure S1).
Input to TEAM are three, respectively six (3 surface, 3 borehole), component waveforms at 100 Hz sampling rate from multiple stations and the corresponding station coordinates.
Furthermore, the model is provided with a set of output locations, at which the PGA should be predicted.
These can be anywhere within the spatial domain of the model and need not be identical with station locations in the training set.

TEAM extracts features from input waveforms using a convolutional neural network (CNN).
The feature extraction is applied separately to each station, but is identical for all stations.
CNNs are well established for feature extraction from seismic waveforms, as they are able to recognize complex features independent of their position in the trace.
On the other hand, CNN based feature extraction usually requires a fixed input length, inhibiting real-time processing.
We allow real-time processing through the alignment of the waveforms and zero-padding: we align all input waveforms in time i.e., all start at the same time $t_0$ and end at the same time $t_1$.
We define $t_0$ to be 5~s before the first P wave arrival at any station, allowing the model to understand the noise characteristics.
For $t_1$ we use the current time, i.e., the amount of available waveforms.
We obtain constant length input, by padding all waveforms after $t_1$ with zeros up to a total length of 30~s.
The feature extraction is described in more detail in supplementary text S2.

TEAM combines the feature vectors and maps them to representations at the targets using a transformer \citep{vaswani2017attention}.
Transformers are attention-based neural networks for combining information from a flexible number of input vectors in a learnable way.
To encode the location of the recording stations as well as of the prediction targets, we use sinusoidal vector representations.
For input stations, we add these representations component-wise to the feature vectors, for target stations we directly use them as inputs to the transformer.
This architecture, processing a varying number of inputs, together with the explicitly encoded locations, allows TEAM to handle dynamically varying sets of stations and targets.
The transformer returns one vector for each target representing predictions at this target.
Details on the feature combinations can be found in supplementary text S3.

From each of the vectors returned by the transformer, TEAM calculates the PGA predictions at one target.
Similar to the feature extraction, the PGA prediction network is applied separately to each target, but is identical for all targets.
TEAM uses mixture density networks \citep{bishop1994mixture} returning Gaussian mixtures, to computes PGA densities.
Gaussian mixtures allow TEAM to predict more complex distributions and better capture realistic uncertainties than a point estimate or a single Gaussian.
The full specifications for the final PGA estimation are provided in supplementary text S4.

TEAM is trained end-to-end using a negative log-likelihood loss.
To increase the flexibility of TEAM and allow for real-time processing, we use training data augmentation.
We randomly select the stations used as inputs and targets in each training iteration.
In addition, again in each training iteration, we randomly replace all waveforms after a time $t$ with zeros, matching the input representation of real time data, to train TEAM for real-time application.
These data augmentations as well as the complete training procedure are further described in supplementary text S5.

To mitigate the systematic underestimation of high PGA values observed in previous machine learning models, TEAM oversamples large events and PGA targets close to the epicenter during training, which reduces the inherent bias in data towards smaller PGAs.
When learning from small catalogs or when applied to regions where events substantially larger than all training events can be expected, e.g., because of known locked fault patches or historic records, TEAM additionally can use domain adaptation.
To this end the training procedure is modified to include large events from other regions that are similar to the expected events in the target region.
While records from those events will differ in certain aspects, e.g., site responses or the exact propagation patterns, other aspects, e.g., the average extent of strong shaking or the duration of events of a certain size, will mostly be independent of the region in question.
The domain adaptation aims to enable the model to transfer the region immanent aspects of large events, at the cost of a certain blurring of the specific regional aspects of the target region.
TEAM aims to mitigate the blurring of regional aspects by the choice of training procedure.

Our Italy dataset is an example of this situation.
Accordingly, TEAM applies domain adaptation to this case: It first trains a joint model using data from Japan and from Italy, which is then fine-tuned using the Italy data on its own, except for the addition of a few large, shallow, onshore events from Japan.
We chose these events, as for Italy one also expects large, shallow, crustal events due to its tectonic setting and earthquake history.
As we use events from Italy in both training steps and in particular in the second step the overwhelming number of events are from Italy, we expect that this scheme only results in a small degradation in the modelling of the regional specifics of the Italy region.

\subsection{Baselines}
We compare TEAM to two state-of-the-art early warning methods, one using source estimation and one propagation based.
As source estimation based method we use the estimated point source approach (EPS), which estimates magnitudes from peak displacement during the P-wave onset \citep{kuyukGlobalApproachProvide2013} and then applies a GMPE \citep{cuaCharacterizingAverageProperties2009} to predict the PGA.
For simplicity, our implementation assumes knowledge of the final catalog epicentre, which is impossible in real-time, leading to overoptimistic results for EPS.
As propagation based method we chose an adaptation of PLUM \citep{koderaPropagationLocalUndamped2018}, which issues warnings if a station within a radius $r$ of the target exceeds the level of shaking.
In contrast to the original PLUM, which operates on the Japanese seismic intensity scale, $I_{JMA}$ \citep{shabestari2001proposal}, our adaptation applies the concept of PLUM to PGA, thereby making it comparable to the other approaches.
Whereas $I_{JMA}$ is also a measure of the strongest acceleration and is thus strongly correlated with PGA, it considers a narrower frequency band and imposes a mimum duration of strong shaking. As such, although the perfomance might vary slightly for our PLUM-like approach compared to the original PLUM, it still exhibits its key features, in particular the effects of the localized warning strategy.
Additionally we apply the GMPE used in EPS to catalog location and magnitude as an approximate upper accuracy bound for point source algorithms (Catalog-GMPE or C-GMPE).
C-CMPE is a theoretical bound that can not be realized in real-time.
It can be considered as an estimate of the modeling error for point source approaches.
A detailed description of the baseline methods can be found in supplementary text S6.

\section{Results}

\subsection{Alert performance}

We compare the alert performance of all methods for PGA thresholds from light (1\%g) to very strong (20\%g) shaking, regarding \emph{precision}, the fraction of alerts actually exceeding the PGA threshold, and \emph{recall}, the fraction of issued alerts among all cases where the PGA threshold was exceeded \citep{meierHowGoodAre2017,minsonLimitsEarthquakeEarly2019}.
Precision and recall trade-off against each other depending on $\alpha$.
While the PGA predictions of TEAM, EPS and the C-GMPE are probabilistic, the thresholding transforms the predictions into alerts or non-alerts.
The probability distribution describes the uncertainty of the models, e.g., for the GMPE the apparent aleatoric uncertainty from aspects not accounted for, which makes false and missed alerts inevitable.
The threshold value controls the trade-off between both types of errors, and its appropriate value will depend on user needs, specifically the costs associated with false and missed alerts.
Therefore, to analyze the performance of the models across different user requirements, we look at the precision recall curves for different thresholds $\alpha$.
In addition to precision and recall, we use two summary metrics: \emph{F1 score}, the harmonic mean of precision and recall, and \emph{AUC}, the area under the precision recall curve.
The evaluation metrics and full setup of the evaluation are defined in detail in supplement text S7.

\begin{figure}
 \includegraphics[width=\textwidth]{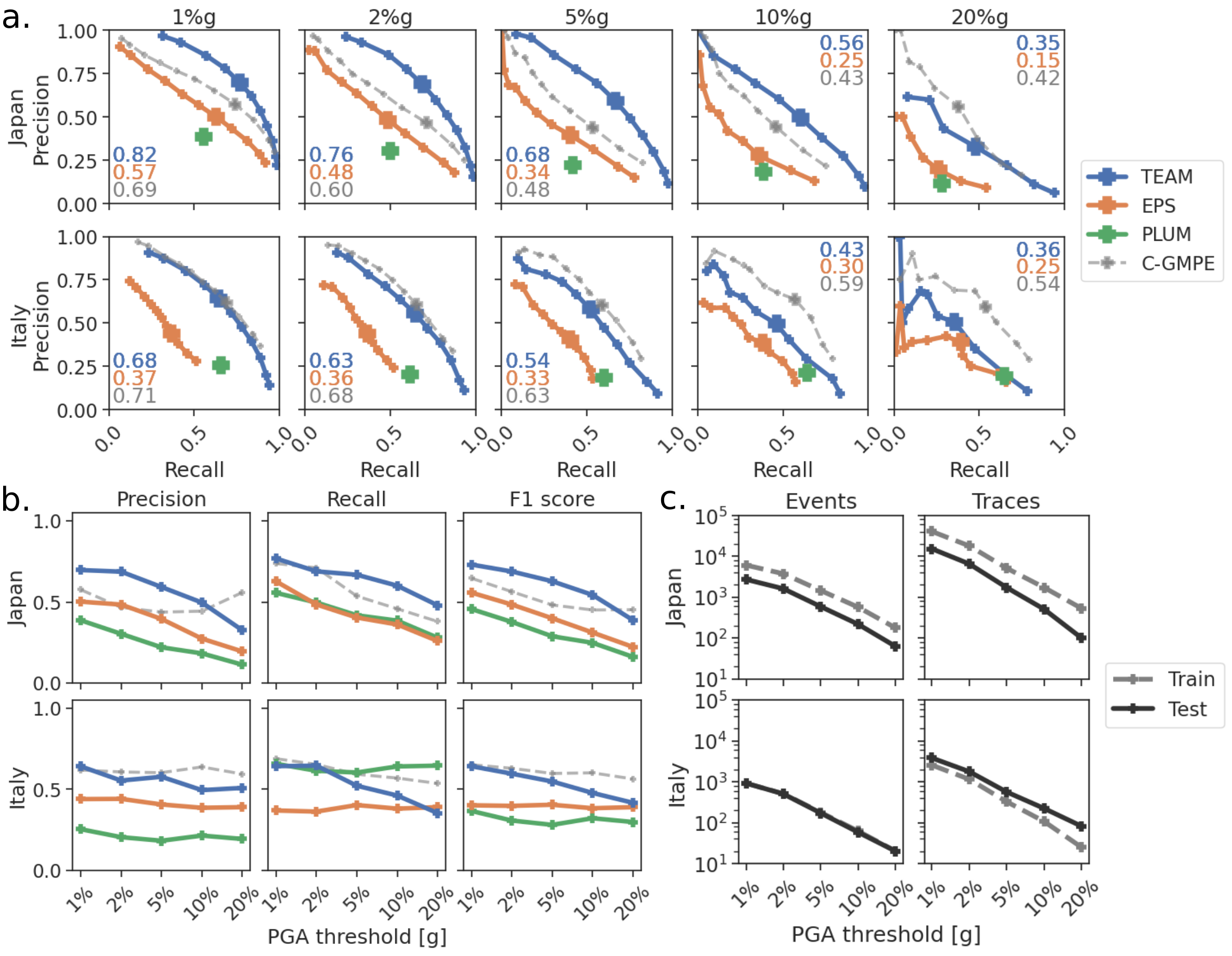}
 \caption{Warning statistics for the three early-warning models (TEAM, EPS, PLUM) for the Japan and Italy datasets. In addition, statistics are provided for C-GMPE, which can only be evaluated post-event due to its reliance on catalog magnitude and location. 
 \textbf{a.} Precision and recall curves across different thresholds $\alpha = 0.05, 0.1, 0.2, \dots, 0.8, 0.9, 0.95$. As the PLUM-like approach has no tuning parameter, its performance is shown as a point. Enlarged markers show the configurations yielding the highest F1 scores. Numbers in the corner give the area under the precision recall curve (AUC), a standard measure quantifying the predictive performance across thresholds.
 \textbf{b.} Precision, recall and F1 score at different PGA thresholds using the F1 optimal value $\alpha$. Threshold probabilities $\alpha$ were chosen independently for each method and PGA threshold.
 \textbf{c.} Number of events and traces exceeding each PGA threshold for training and test set. Training set numbers include development events and show the numbers before oversampling is applied. 
 For Italy training and test event curve are overlapping due to similar numbers of events.}
 \label{fig:results}
\end{figure}

TEAM outperforms both EPS and the PLUM-like approach for both datasets and all PGA thresholds, indicated by the precision-recall-curves of TEAM lying to the top-right of the baseline curves (Figure \ref{fig:results}a).
For any baseline method configuration, there is a TEAM configuration surpassing it both in precision and in recall.
Improvements are larger for Japan, but still substantial for Italy.
To compare the performance at fixed $\alpha$, we selected $\alpha$ values yielding the highest F1 score separately for each PGA threshold and method.
Again, TEAM outperforms both baselines on both datasets, irrespective of the PGA level (Figure \ref{fig:results}b).
Performance statistics in numerical form are available in tables S1 and S2.

All methods degrade with increasing PGA levels, particularly for Japan.
This degradation is intrinsic to early warning for high thresholds due to the very low prior probability of strong shaking \citep{meierHowOftenCan2020,minsonLimitsEarthquakeEarly2019,meierHowGoodAre2017}.
Furthermore, shortage of training data with high PGA values results in less well constrained model parameters.

Using domain adaptation techniques, TEAM copes well with the Italy data, even though the largest test event ($M_w = 6.5$) is significantly larger than the largest training event ($M_w = 6.1$), and three further test events have $M_W \geq 5.8$.
To assess the impact of this technique, we compared TEAM's results to a model trained without it (Figures S2, S3).
While for low PGA thresholds differences are small, at high PGA levels they grow to more than 20 points F1 score.
Interestingly, for large events, TEAM strongly outperforms TEAM without domain adaptation even for low PGA thresholds.
This shows that domain adaptation does not only allow the model to predict higher PGA values, but also to accurately assess the region of lighter shaking for large events.
Domain adaptation therefore helps TEAM to remain accurate even for events quite far from the training distribution.

\subsection{Warning times}
\begin{figure}
 \includegraphics[width=\textwidth]{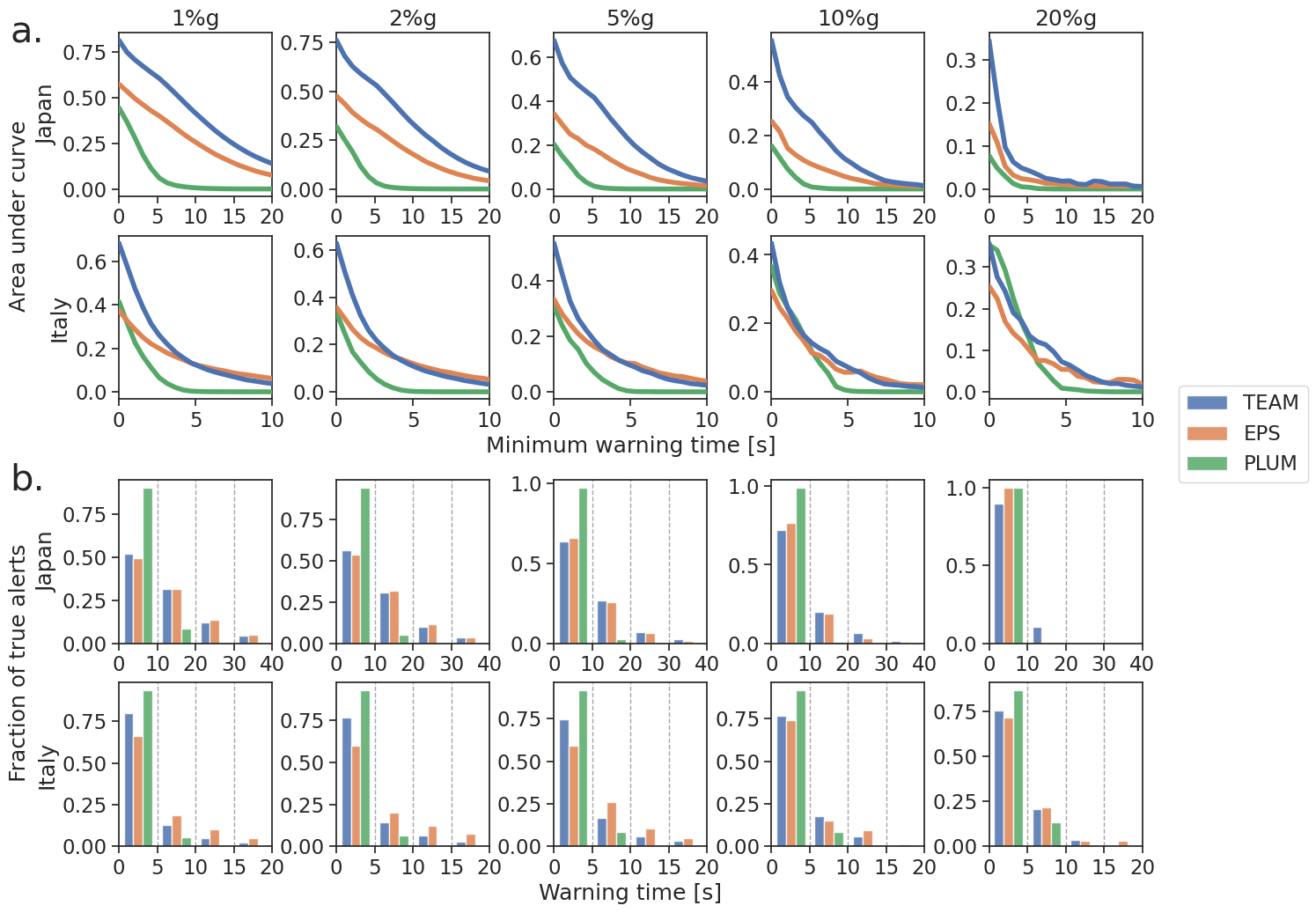} 
 \caption{Warning time statistics.
 \textbf{a.} Area under the precision recall curve for different minimum warning times. All alerts with shorter warning times are counted as false negatives.
 \textbf{b.} Warning time histograms showing the distribution true alerts across distances for the different methods.
 Please note that the total number of true alerts differs by method and is not shown in this subplot.
 Therefore the values of different methods can not be directly compared, but only the differences in the distributions.
 TEAM and EPS are shown at F1-optimal $\alpha$, chosen separately for each threshold and method.
 Warning time dependence on hypocentral distance is shown in Figure S4.}
 \label{fig:warning_times}
\end{figure}

In application scenarios, a user will usually require a certain warning time, which is the time between issuing of the warning and first exceedance of the level of shaking, as this time is necessary for taking action.
As the previous evaluation considered prediction accuracy irrespective of the warning time, we now compare the methods while imposing a certain minimum warning time.
Actually, TEAM consistently outperforms both baselines across different required warning times and irrespective of the PGA threshold (Figure \ref{fig:warning_times}a).
While the margin for TEAM compared to the baselines is smaller for Italy than for Japan, TEAM shows consistently strong performance across different warning times.
In contrast, EPS performs clearly worse at short warning times, the PLUM-based approach at longer warning times.
The latter is inherent to the key idea of PLUM and makes the method only competitive at high PGA thresholds, where potential maximum warning times are naturally short due to the proximity between stations with strong shaking and the epicenter \citep{minsonLimitsEarthquakeEarly2018}.
We further note that while the PLUM-like approach shows slightly higher AUC than TEAM for short warning times at 20~\%g, this is only a hypothetical result.
As PLUM does not have a tuning parameter between precision and recall, this performance can actually only be realised for a specific precision/recall threshold, where it performs slightly superior to TEAM (Figure \ref{fig:results}a bottom right).

Warning times depend on $\alpha$: a lower $\alpha$ value naturally leads to longer warning times but also to more false positive warnings.
At F1-optimal thresholds $\alpha$, EPS and TEAM have similar warning time distributions (Figure \ref{fig:warning_times}b, Table S3), but lowering $\alpha$ leads to stronger increases in warning times for TEAM.
For instance, at 10\%g, lowering $\alpha$ from 0.5 to 0.2 increases average warning times of TEAM by 2.3~s/1.2~s (Japan/Italy), but only by 1.1~s/0.1~s for EPS.
Short times as measured here are critical in real applications: First, they reduce the time available for counter measures.
Second, real warning times will be shorter than reported here due to telemetry and compute delays.
However, compute delays for TEAM are very mild: analysing the Norcia event (25 input stations, 246 target sites) for one time step took only 0.15~s on a standard workstation using non-optimized code.

\section{Discussion}

\subsection{Calibration of probabilities}

Even though TEAM and EPS give probabilistic predictions, it is not clear whether these predictions are well-calibrated, i.e., if the predicted confidence values actually correspond to observed probabilities.
Calibrated probabilities are essential for threshold selection, as they are required to balance expected costs of taking action versus expected costs of not taking action.
We note that while good calibration is a necessary condition for a good model, it is not sufficient, as a model constantly predicting the marginal distribution of the labels would be always perfectly calibrated, yet not very useful.

To assess the calibration, we use calibration diagrams (Figures S9 and S10) for Japan and Italy at different times after the first P arrival.
These diagrams compare the predicted probabilities to the actually observed fraction of occurrences.
In general, both models are well calibrated, with a slightly better calibration for TEAM.
Calibration is generally better for Japan, where only EPS is slightly underconfident at earlier times for the highest PGA thresholds.
For Italy, EPS is generally slightly overconfident, while TEAM is well calibrated, except for a certain overconfidence at $20\%$g.
We suspect that the worse calibration for the largest events is caused by the domain adaptation strategy, but the better performance in terms of accuracy clearly weighs out this downside of domain adaptation.

\subsection{Insights into the transformer earthquake alerting model}

\begin{figure}
 \includegraphics[width=\textwidth]{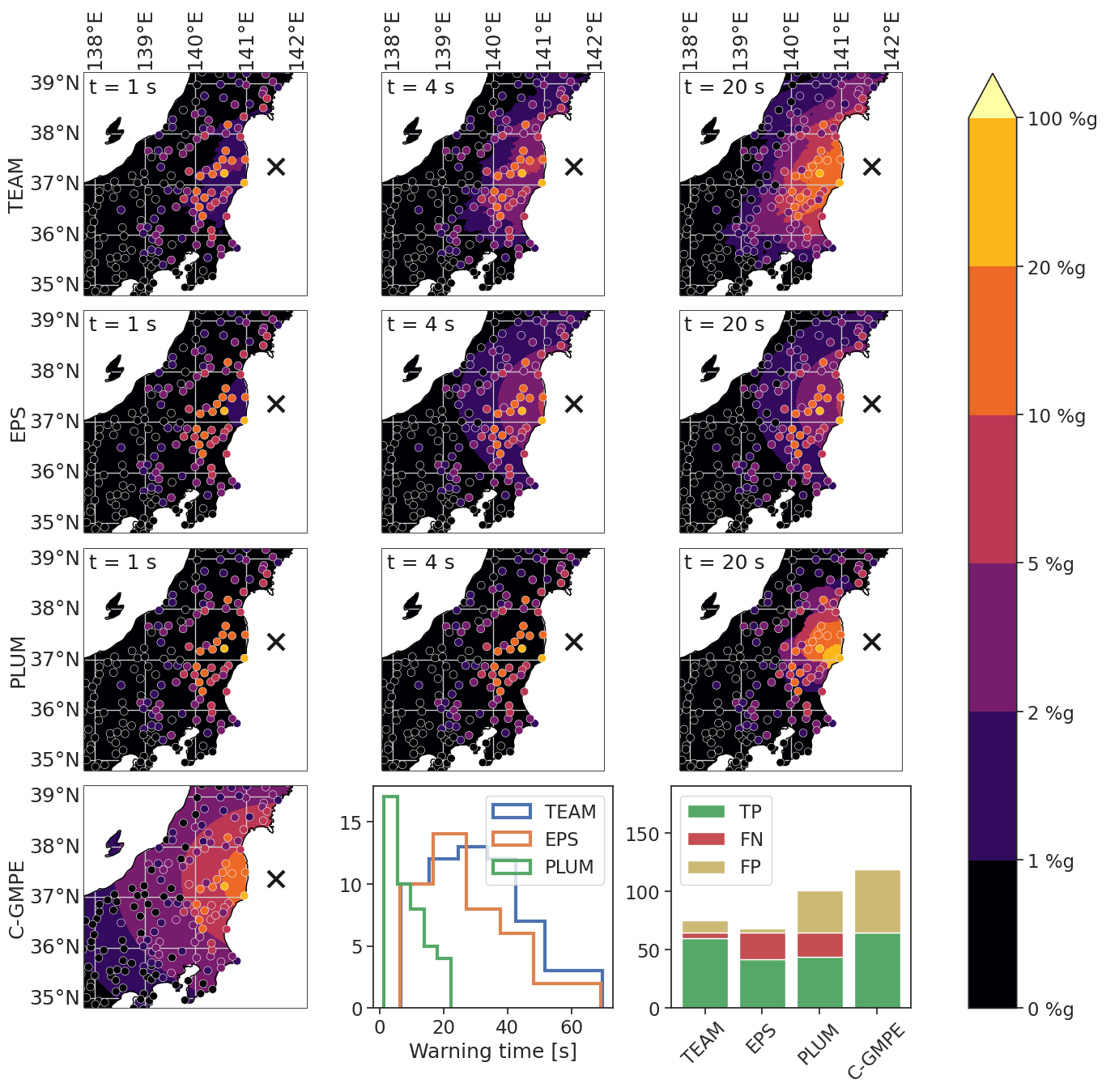}
 \caption{Scenario analysis of the 22nd November 2016 $M_J=7.4$ Fukushima earthquake, the largest test event located close to shore. Maps show the warning levels for each method (top three rows) at different times (shown in the corner, $t=0$ s corresponds to P arrival at closest station). Dots represent stations and are colored according to the PGA recorded during the full event, i.e., the prediction target. The bottom row shows (left to right), the catalog based GMPE predictions, the warning time distributions, and the true positives (TP), false negatives (FN) and false positives (FP) for each method, both at a 2\%g PGA threshold.
 EPS and GMPE shake map predictions do not include station terms, but they are included for the bottom row histograms.}
 \label{fig:scenario}
\end{figure}

We analyze differences between the methods using one example event from each dataset (Japan: Figure \ref{fig:scenario}, Italy: Figure S5).
All methods underestimate the shaking in the first seconds (left column Figures \ref{fig:scenario}, S5).
However, TEAM is the quickest to detect the correct extent of the shaking.
Additionally, it estimates even fine-grained regional shaking details in real-time (middle and right columns).
In contrast, shake maps for EPS remain overly simplified due to the assumptions inherent to GMPEs (right column and bottom left panel).
For the Japan example, even late predictions of EPS understimate the shaking, due to an underestimation of the magnitude.
The PLUM-based approach produces very good PGA estimates, but exhibits the worst warning times.

Notably, TEAM predictions at later times correspond even better to the measured PGA than C-GMPE estimates, although these are based on the final magnitude (top right and bottom left panels).
For the Japan data, this is not only the case for the example at hand, but also visible in Figure \ref{fig:results}, showing higher accuracy of TEAM's prediction compared to C-GMPE for all thresholds except 20\%g on the full Japan dataset.
We assume TEAM's superior performance is rooted in both global and local aspects.
Global aspects are the abilities to exploit variations in the waveforms, e.g., frequency content, to model complex event characteristics, such as stress drop, radiation pattern or directivity, and to compare to events in the training set.
Local aspects include understanding regional effects, e.g., frequency dependent site responses, and the ability to consider shaking at proximal stations.
We note that for our Italy experiments, the modelling of local aspects resulting from regional characteristics might be slightly degraded by the domain adaptation.
However, the first-order propagation effects such as, e.g., amplitude decay due to geometric spreading, are similar between regions and therefore not negatively affected by the domain adaptation.
In conclusion, combining a global event view with propagation aspects, TEAM can be seen as a hybrid model between source estimation and propagation.

\subsection{TEAM performance on the Tohoku sequence}

\begin{figure}
 \includegraphics[width=\textwidth]{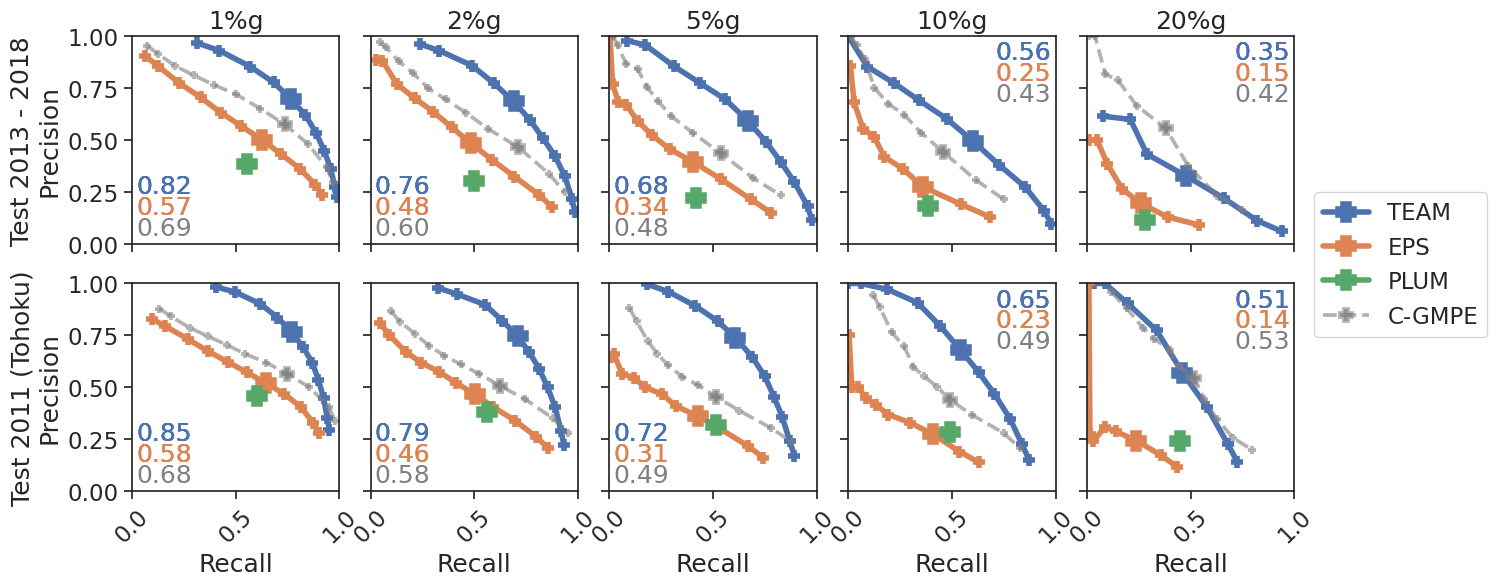}
 \caption{Precision recall curves for the Japanese dataset using the chronological split (top) and using the events in 2011 as test set (bottom). The year 2011 contains the $M_w=9.1$ Tohoku event as well as its aftershocks.}
 \label{fig:prec_rec_tohoku}
\end{figure}

We evaluated TEAM for Japan on a chronological train/dev/test split, as this split ensures the evaluation closest to the actual application scenario.
On the other hand, this split put the $M=9.1$ Tohoku event in March 2011 into the training set.
To evaluate the performance for this very large event and its aftershocks, we trained another TEAM instance using the year 2011 as test set and the remainder of the data for training and validation.
Figure \ref{fig:prec_rec_tohoku} shows the precision recall curves for the chronological split and the year 2011 as test set.
In general, the performance of all models stays similar when evaluated on the alternative split.
A key difference between the curves is, that TEAM, in particular for high PGA thresholds, does not reach similar levels of recall for 2011 as for the chronological split, while achieving higher precision. As we will describe in the next paragraph, this trend probably results from a tendency to underestimate true PGA amplitudes, which will naturally reduce recall and boost precision.
Nevertheless, the performance of TEAM as quantified by the AUC actually improves, and significantly so for the highest thresholds.
We suspect that this tendency for underestimation is either caused by the higher number of large events in the 2011 test set compared to the chronological split, or by the lower number of high PGA events in the training set without 2011.

Figure S6 presents a scenario analysis for the Tohoku event.
All models underestimate the event considerably, with the strongest underestimation for the EPS method.
Even 20~s after the first P wave arrival, all methods underestimate both the severity and the extent of shaking.
Do to its localized approach, the PLUM-based model achieves the highest number of true warnings, albeit at short warning times and a certain number of false positives, which due to the underestimation are totally absent from TEAM and EPS predictions.
The performance of both EPS and TEAM is likely degraded by the slow onset of the Tohoku event as described by \citep{koketsu2011unified}.
According to \citep{koketsu2011unified} the main subevent with a displacement of 36~m only initiated 20~s after the onset of the Tohoku event.
Therefore only the first P waves for EPS or at most the first 25~s of waveforms for TEAM is most likely insufficient to correctly estimate the size of the Tohoku event.

For Italy, we showed that underestimation for large events can be mitigated using transfer learning.
However, the Tohoku event clearly shows the limitations of this strategy, as nearly no training data for events of comparable size are available, even when using events across the globe.
Therefore, for the largest events alternative strategies need to be developed, e.g., training using simulated data.
Furthermore, the 25~s of waveforms used by TEAM in the current implementation may, for a very large event, not capture the largest subevent.
While we decided to use only 25~s of event waveforms, as there is only insufficient training data of longer events, this window could be extended when developing training strategies and models for the largest events.

\section{Conclusion}

In this study we presented the transformer earthquake alerting model (TEAM).
TEAM outperforms existing early warning methods in terms of both alert performance and warning time.
Using a flexible machine learning model, TEAM is able to extract information about an event from raw waveforms and leverage the information to model the complex dependencies of ground motion.
We point out two further aspects that make TEAM appealing to users.
First, TEAM can adapt to various user requirements by combining two thresholds, one for shake level and one for the exceedance probability.
As TEAM outputs probability density functions over the PGA, these thresholds can easily be adjusted by individual users on the fly, e.g., by setting sliders in an early warning system.
Second, deep learning models typically exhibit large performance improvements from larger training datasets \citep{sun2017revisiting} due to the high number of model parameters.
In our study this reflects in the better performance on the twofold larger Japan dataset.
This indicates that TEAM's performance can be improved just by collecting more comprehensive catalogs, which happens automatically over time.

\section*{Acknowledgements}
We thank the National Research Institute for Earth Science and Disaster Resilience (NIED) for providing the catalog and waveform data for our Japan dataset.
We thank the Istituto Nazionale di Geofisica e Vulcanologia and the Dipartimento della Protezione Civile for providing the catalog and waveform data for our Italy dataset.
J. M. acknowledges the support of the Helmholtz Einstein International Berlin Research School in Data Science (HEIBRiDS).
We thank Matteo Picozzi for discussions on earthquake early warning that helped improve the study design.
We thank Hiroyuki Goto and an anonymous reviewer for their comments which helped to improve the manuscript.
An implementation of TEAM and the baselines is available at \url{https://github.com/yetinam/TEAM}.
The Italy dataset has been published as \cite{munchmeyer2020dataitaly}.
The Japan dataset can be obtained using the scripts in the code repository.

\nocite{vaswani2017attention,bishop1994mixture,snoek2019can,cuaCharacterizingAverageProperties2009,meierHowGoodAre2017,cuaCharacterizingAverageProperties2009,minsonLimitsEarthquakeEarly2019,boseFinDerImprovedRealtime2018,meierHowOftenCan2020,kuyukGlobalApproachProvide2013,koderaPropagationLocalUndamped2018,karimCorrelationJMAInstrumental2002,cochranEventDetectionPerformance2019,nied2019kiknet,network_3a,network_ba,network_fr,network_gu,network_it,network_iv,network_ix,network_mn,network_ni,network_ox,network_ra,network_st,network_tv,network_xo,waldRelationshipsPeakGround1999}

\bibliographystyle{unsrt}

\clearpage
\appendix

\section{Data and Preprocessing}

For our study we use two datasets, one from Japan, one from Italy.
The Japan dataset consists of 13,512~events between 1997 and 2018 from the NIED KiK-net catalog \citep{nied2019kiknet}.
The data was obtained from NIED and consists of triggered strong motion records.
Each trace contains 15~s of data before the trigger and has a total length of 120 s.
Each station consists of two three component strong motion sensors, one at the surface and one borehole sensor.
We split the dataset chronologically with ratios of 60:10:30 between training, development and test set.
The training set ends in March 2012, the test set begins in August 2013.
Events in between are used as development set.
We decided to use a chronological split to ensure a scenario most similar to the actual application in an early warning setting.

The Italy dataset consists of 7,055~events between 2008 and 2019 from the INGV catalog.
We use data from the 3A \citep{network_3a}, BA \citep{network_ba}, FR \citep{network_fr}, GU \citep{network_gu}, IT \citep{network_it}, IV \citep{network_iv}, IX \citep{network_ix}, MN \citep{network_mn}, NI \citep{network_ni}, OX \citep{network_ox}, RA \citep{network_ra}, ST \citep{network_st}, TV \citep{network_tv} and XO \citep{network_xo} networks.
We use all events from 2016 as test set and the remaining events as training and development sets.
The test set consists of 31\% of the events, a similar fraction as in the Japan dataset.
We shuffle events between training and development set.
While a chronological split would have been the default choice, we decided to  use 2016 for testing, as it contains a long seismic sequence in central Italy containing several very large events in August and October. 
Further details on the statistics of both datasets can be found in Table \ref{tab:datasets}.

Before training we extract, align and preprocess the waveforms and store them in hdf5 format.
As alignment requires the first P pick, we need approximate picks for the datasets.
For Japan we use the trigger times provided by NIED.
Our preprocessing accounts for misassociated triggers.
For Italy we use an STA/LTA trigger around the predicted P arrival.
While triggering needs to be handled differently in an application scenario, we use this simplified approach as our evaluation metrics depend only very weakly on the precision of the picks.

\section{TEAM - Feature extraction network}

The feature extraction of TEAM is conducted separately for each station.
Nonetheless the same convolutional neural network (CNN) for feature extraction is applied at all stations, i.e., the same model with the same model weights.

As amplitudes of seismic waveforms can span several orders of magnitude, the first layer of the network normalizes the traces by dividing through their peak value observed so far.
All components of one station are normalized jointly, such that the amplitude ratio between the components stays unaltered.
Notably, we only use the peak value observed so far, i.e., the waveforms after $t_1$, which have been blinded with zeros, are not considered, as this would introduce a knowledge leak.
As the peak amplitude of the trace is likely a key predictor, we logarithmize the value and concatenate it to the feature vector after passing through all the convolutional layers, prior to the fully connected layers.

We apply a set of convolutional and max-pooling layers to the waveforms.
We use convolutional layers as this allows the model to extract translation invariant features and as convolutional kernels can be interpreted as modeling frequency features.
We concatenate the output of the convolutions and the logarithm of the peak amplitude.
This vector is fed into a multi-layer perceptron to generate the final features vector for the station.
All layers use ReLu activations.
A detailed overview of the number and specifications of the layers in the feature extraction model can be found in Table \ref{tab:convolutions}.

\section{TEAM - Feature combination network}

The feature extraction provides one feature vector per input station representing the waveforms.
As additional input the model is provided with the location of the stations, represented by latitude, longitude and elevation.
The targets for the PGA estimation are specified by the latitude, longitude and elevation as well.

We use a transformer network \citep{vaswani2017attention} for the feature combination.
Given a set of $n$ input vectors, a transformer produces $n$ output vectors capturing combined information from all the vectors in a learnable way.
We use transformers for two main reasons.
First, they are permutation equivariant, i.e., changing the order of input or output stations does not have any impact on the output.
This is essential, as there exists no natural ordering on the input stations.
Second, they can handle variable input sizes, as the number of parameters of transformers is independent of the number of input vectors.
This property allows application of the model to different sets of stations and a flexible number of target locations.

To incorporate the locations of the stations we use predefined position embeddings.
As proposed by \citep{vaswani2017attention}, we use pairs of sinusoidal functions, $\sin(\frac{2\pi}{\lambda_i} x)$ and $\cos(\frac{2\pi}{\lambda_i} x)$, with different wavelength $\lambda_i$.
We use 200 dimensions for latitude and longitude, respectively, and the remaining 100 dimensions for elevation.
We anticipate two advantages of sinusoidal embeddings for representing the station position.
First, keeping the position embeddings fixed instead of learnable reduces the parameters and therefore likely provides better representations for stations with only few input measurements or sites not contained in the training set.
Second, sinusoidal embeddings guarantee that shifts can be represented by linear transformations, independent of the location it applies to.
As the attention mechanism in transformers is built on linear projections and dot products, this should allow for more efficient attention scores at least in the first transformer layers.
As proposed in the original transformer paper \citep{vaswani2017attention}, the position embeddings are added element-wise to the feature vectors to form the input of the transformer.
We calculate position embeddings of the target locations in the same way.

As in our model input and output size of the transformer are identical, we only use the transformer encoder stack \citep{vaswani2017attention} with six encoder layers.
Inputs are the feature vectors with position embeddings from all input stations and the position embeddings of the output locations.
By applying masking to the attention we ensure that no attention weight is put on the vectors corresponding to the output locations.
This guarantees that each target only affects its own PGA value and not any other PGA values.
As the self-attention mechanism of the transformer has quadratic computational complexity in the number of inputs, we restrict the maximum number of input stations to 25  (see training details for the selection procedure).
Further details on the hyperparameters can be found in Table \ref{tab:transformer}.
The transformer returns one output vector for each input vector.
We discard the vectors corresponding to the input stations and only keep the vectors corresponding to the targets.

\section{TEAM - Mixture density output}

Similar to the feature extraction, the output calculation is conducted separately for each target, while sharing the same model and weights between all targets.
We use a mixture density network to predict probability densities for the PGA \citep{bishop1994mixture}.
We model the probability as a mixture of $m=5$ Gaussian random variables.
Using a mixture of Gaussians instead of a single Gaussian allows the model to predict more complex distribution, like non-Gaussian distributions, e.g., asymmetric distributions.
The functional form of the Gaussian mixture is $\sum_{i=1}^{m} \alpha_i \varphi_{\mu_i, \sigma_i}(x)$.
We write $\varphi_{\mu_i, \sigma_i}$ for the density of a standard normal with mean $\mu_i$ and standard deviation $\sigma_i$.
The values $\alpha_i$ are non-negative weights for the different Gaussians with the property $\sum_{i=1}^m \alpha_i = 1$.
The mixture density network uses a multi-layer percepton to predict the parameters $\alpha_i$, $\mu_i$ and $\sigma_i$.
The hidden dimensions are 150, 100, 50, 30, 10.
The activation function is ReLu for the hidden layers, linear for the $\mu$ and $\sigma$ outputs and softmax for the $\alpha$ output.

\section{TEAM - Training details}

We train the model end-to-end using negative log-likelihood as loss function.
All components are trained jointly end-to-end.
The model has about 13.3 million parameters in total.
To increase the amount of training data and to train the model on shorter segments of data we apply various forms of data augmentation.
Each data augmentation is calculated separately each time a particular waveform sample is shown, such that the effective training samples vary.

First, if our dataset contains more stations for an event than the maximum number of 25 allowed by the model, we subsample.
We introduce a bias to the subsampling to favor stations closer to the event.
We use up to twenty targets for PGA prediction.
Similarly to the input station, we subsample if more targets are available and bias the subsampling to stations close to the event.
This bias ensures that targets with higher PGA values are shown more often during training.

Second, we apply station blinding, meaning we zero out a set of stations in terms of both waveforms and coordinates.
The number of stations to blind is uniformly distributed between zero and the total number of stations available minus one.
In combination with the first point this guarantees that the model also learns to predict PGA values at sites where no waveform inputs are available.

Third, we apply temporal blinding.
We uniformly select a time $t$ that is between 1~$s$ before the first P pick and 25~$s$ after.
All waveforms are set to zero after time $t$.
The model therefore only uses data available at time $t$.
Even though we never apply TEAM to times before the first P pick, we include these in the training process to ensure TEAM learns a sensible prior distribution.
We observed that this leads to better early predictions.
As information about the triggering station distribution would introduce a knowledge leak if available from the beginning, we zero out all waveforms and coordinates from stations that did not trigger until time $t$.

Fourth, we oversample large magnitude events.
As large magnitude events are rare, we artificially increase their number in the training set.
An event with magnitude $M \geq M_0$ is used $\lambda^{M - M_0}$ times in each training epoch with $\lambda = 1.5$ and $M_0 = 5$ for Japan and $M_0 = 4$ for Italy.
This event-based oversampling implicitly increases the number of high PGA values in the training set, too.

We apply all data augmentation on the training and the development set, to ensure that the development set properly represents the task we are interested in.
As this introduces stochasticity into the development set metrics, we evaluate the development set three times after each epoch and average the result.
In contrast, at test time we do not apply any data augmentation, except temporal blinding for modelling real-time application.
If more than 25 stations are available for a test set event, we select the 25 station with the earliest arrivals for evaluation.

We train our model using the Adam optimizer.
We emphasize that the model is only trained on predicting the PGA probability density and does not use any information on the PGA thresholds used for evaluation.
We start with a learning rate of $10^{-4}$ and decrease the learning rate by a factor of 3 after 5 epochs without a decrease in validation loss.
For the final evaluation we use the model from the epoch with lowest loss on the development set.
We apply gradient clipping with a value of 1.0.
We use a batch size of 64.
We train the model for 100 epochs.

To improve the calibration of the predicted probability densities we use ensembles \citep{snoek2019can}.
We use an ensemble size of 10 models and average the predicted probability densities.
We weight each ensemble member identically.
To increase the entropy between the ensembles, we also modify the position encodings between the ensemble members by rotating the latitude and longitude values of stations and targets.
The rotations for the 10 ensemble members are $0^\circ, 5^\circ, \dots, 40^\circ, 45^\circ$.

For the Italy model we use domain adaptation by modifying the training procedure.
We first train a model jointly on the Italy and Japan data set, according to the configuration described above.
We use the resulting model weights as initialization for the Italy model.
For this training we reduce the number of PGA targets to 4, leading to a higher fraction of high PGA values in the training data, and the learning rate to $10^{-5}$.
In addition, we train jointly on an auxilary data set, comprised of 77 events from Japan.
The events were chosen to be shallow, crustal and onshore, having a magnitude between 5.0 and 7.2.
We shift the coordinates of the stations to lie in Italy.
We use 85\% of the auxiliary events in the training set and 15\% in the development set.

We implemented the model using Tensorflow.
We trained each model on one GeForce RTX 2080 Ti or Tesla V100.
Training of a single model takes approximately 5~h for the Japan dataset, 10~h for the joint model and 1~h for the Italy data set.
We benchmarked the inference performance of TEAM on a common workstation with GPU acceleration (Intel i7-7700, Nvidia Quadro P2000).
Running TEAM with ensembling at a single timestep took 0.15~s for all 246 PGA targets of the Norcia event.
As our implementation is not optimized for run time, we expect an optimized implementation to yield multifold lower run times, enabling a real-time application of TEAM with high update rate and low compute latency.

Figure \ref{fig:loss_curves} shows the training and validation loss curves for the Japan TEAM model and the fine-tuning step of the Italy TEAM model.
While there is some variation between the ensemble members, all show similar characteristics.
We note, that the early appearance of the optima for the Italy fine-tuning is expected, because of the transfer learning applied.
We validated through a comparison of the fine-tuned and the non-fine-tuned models, that the fine-tuning step still leads to considerable improvement in the model performance.

As visible from the by far lower training than validation loss, all models exhibit overfitting.
This is expected, as the number of model parameters (13.3M) is very high in comparison to the number of training examples (<10,000).
However, multiple publications (e.g. \cite{belkin2019reconciling,muthukumar2020harmless}) have provided theoretical and empirical evidence, that overfitting for deep learning is not problematic and can even lead to considerably better performance than a not overfitted model, if proper model selection on a validation set is employed.
We conduct this model selection by using the model with the lowest validation score.

\section{Baseline methods}

We compare TEAM to two baseline methods, EPS and a PLUM-based approach.
We do not compare to any deep learning baseline, because we are not aware of any published deep learning method for early warning that can actually be applied in real-time.
For the EPS method we use a GMPE based on the functional form by  \citep{cuaCharacterizingAverageProperties2009} and add a quadratic magnitude term as proposed by  \citep{meierHowGoodAre2017}.
We make further minor adjustments to accommodate the wider range of magnitudes in our datasets.
The functional form of the GMPE is:
\begin{align}
 \log(pga) &= a_1 M + a_2 \max(M - M_0, 0) ^ 2 + b (R_d + C(M)) + d \log(R_d + C(M)) + e + \delta_S + \mathscr{N}(0, \sigma^2)\\
 C(M) &:= c_1 \exp(c_2 \max(0, M - 5)) (\arctan(M - 5) + \pi / 2)\\
 R_d &:= \sqrt{R^2 + H_d^2}
\end{align}
We write $M$ for magnitude, $R$ for epicentral distance, $\delta_S$ for the station bias, and $e$ for an error term.
We use $m/s^2$ as unit for PGA and $km$ as unit for all length measurements.
We use a pseudo-depth $H_d$, depending on the event depth and the dataset.
This allows to model the stronger attenuation with distance for shallow events.
For Italy we set $H_d = 5$~km for events shallower than 20~km and $H_d = 50$~km for all other events.
For Japan we set $H_d = 5$~km for events shallower than 20~km, $H_d = 40$~km for events between 20~km and 200~km and set $H_d$ to the actual depth for all deeper events, to account for a few very deep events.
We set $M_0 = 4$ for Italy and $M_0 = 6$ for Japan.

We fix $c_1 = 1.48$ and $c_2 = 1.11$, as proposed by \citep{cuaCharacterizingAverageProperties2009}, and optimize the other parameters using linear regression.
We perform the optimization iteratively to obtain station bias terms, using the union of training and development set.
To avoid noise samples in calibration we only use stations for which $R_d < (M - 3.5) * 200\text{~km}$ for Japan and $R_d < (M - 3) * 50\text{~km}$ for Italy.
The calibrated GMPEs have residual values $\sigma$ of 0.29 for Italy and 0.33 for Japan, matching the value of $\sim$0.3 proposed as the approximate current optimum for GMPEs \citep{minsonLimitsEarthquakeEarly2019}.
Residual plots can be found in Figure \ref{fig:residuals}.

We note that our GMPE model is using a point source assumption, which is incorrect for larger events.
We chose this simplification, as it is common in source based early warning and makes the GMPE performance an upper bound for any method relying on magnitude and location estimate.
While there are early warning methods based on extended fault models \citep{boseFinDerImprovedRealtime2018}, they perform equally well as point source approaches for all but the largest events \citep{meierHowOftenCan2020}.
As lower thresholds are dominated by smaller events, for which the point source approximation is valid, the inferior performance of the GMPE compared to TEAM is not an artifact of the point source assumption, but probably related to its inability to account for systematic propagation effects caused by regional structure, and variability of the earthquake source (focal mechanism, stress drop) not captured by the magnitude and location. 

For magnitude estimation we use the peak displacement based method proposed by  \citep{kuyukGlobalApproachProvide2013}.
We bandpass filter the signal between 0.5~Hz and 3~Hz and discard traces with insufficient signal to noise ratio.
We extract peak displacement from the horizontal components in the first 6~s of the P wave.
We stop the time window at latest at the S onset.
We use the relationship
\begin{align}
 M = c_1 \log(PD) + c_2 \log(R) + c_3 + \mathscr{N}(0, \sigma^2)
\end{align}
to estimate magnitudes from peak displacement.
We use $c_1 = 1.23$, $c_2 = 1.38$, $c_3 = 5.69$ (Italy) / $c_3 = 5.89$ (Japan) and $\sigma = 0.31$.
These are the values from \citep{kuyukGlobalApproachProvide2013}, except for $c_3$ which needed to be adjusted as we do not use moment magnitude.
We combine the predictions in probability space assuming independence between the predictions from different stations.
We weight stations based on the length of the P wave window recorded so far.
We use the mean value of the single-station magnitude estimates for PGA estimation.
For both the application of the GMPE and the magnitude estimation we use the catalog hypocenters.
As the quality of real-time location estimates will be worse, this leads to inflated performance measures for EPS.

As second baseline, we adapted the PLUM algorithm \citep{koderaPropagationLocalUndamped2018}.
While the original paper applies PLUM to seismic intensities, we apply it to PGA values.
This adaptation is possible, as approximate linear and especially monotonic relations exist between intensity and PGA \citep{karimCorrelationJMAInstrumental2002}.
However, as seismic intensiy incorporates a narrower frequency band and also considers the duration of strong shaking \citep{shabestari2001proposal}, the PLUM adaptation to PGA might exhibit slightly different performance.
The PGA prediction $\hat{pga}^s_t$ at a station $s$ at time $t$ is the maximum of all observed PGA values $pga^{s'}_t$ at stations $s'$ within a radius $r$ of $s$.
Therefore a warning for a certain threshold for a station is issued once the threshold has been exceeded at any station within the radius $r$.
Due to different station densities in Italy and Japan we used different values for $r$.
For Italy we used $r=15$~km for Japan we used $r=30$~km.
Following the findings of Cochran et al \citep{cochranEventDetectionPerformance2019}, we do not use site correction terms in our implementation of PLUM as they only have minor impact on the performance.

\section{Evaluation metrics}

We analyze the performance of the early warning algorithms using PGA thresholds of 1\%g, 2\%g, 5\%g, 10\%g and 20\%g, approximately matching Modified Mercalli Intensity (MMI) III (light) to VII (very strong) \citep{waldRelationshipsPeakGround1999}.
We calculate PGA from the absolute value of the two horizontal components.
For the PGA values for the Japanese data, we use the surface stations and not the borehole stations.

A warning at a site should be issued if anytime during the event the PGA threshold is exceeded at the site.
We consider a warning correct (true positive, TP), if a warning for a certain threshold was issued and the threshold was actually exceeded later during the event.
Missed warnings (false negative, FN) are all cases, where the PGA threshold was exceeded, but no warning was issued or the warning was issued after the PGA threshold was first exceeded.
We consider a warning false (false positive, FP), if a warning was issued, but the threshold was not exceeded.
All remaining cases are true negatives (TN).

As the number of true negatives depends strongly on the inclusion criteria of the catalog, we use metrics independent of the true negatives.
As summary statistics we use \emph{precision}, TP/(TP+FP), measuring the fraction of correct warnings among all warnings, and \emph{recall}, TP/(TP+FN), measuring the fraction of possible correct warnings that was issued.
We use the \emph{F1 score} = $2 * \text{precision} * \text{recall} / (\text{precision} + \text{recall})$ as a combined statistic.
Any analysis using a fixed $\alpha$ uses the value maximizing the F1 score, which is specific to each method and PGA threshold.
For an analysis independent of the threshold $\alpha$ we use the area under the precision recall curve (AUC).
We use values $\alpha=0.05, 0.1, 0.2, \dots, 0.8, 0.9, 0.95$ and add additional points at $(0, 1)$ and $(1, 0)$ to the precision recall curve to approximate the AUC.
For comparison of the PLUM-based model using AUC in Figure 3, we introduce an artificial precision recall line for PLUM with a slope of $-1$ going through the observed precision and recall values.

We define the warning time as the time between the issuance of a warning and the first exceedance of the threshold.
We consider a zero latency system and do not impose a minimum warning time.
For comparing warning times between methods or different parameter combinations, we only use the subset of station event pairs, where both methods/parameter combinations issued correct warnings.

We evaluate our PLUM-based implementation continuously, i.e., warnings are issued immediately at the exceedance of a threshold.
TEAM and EPS are evaluated every 0.1~s, starting 1~s after the first P arrival for EPS and 0.5~s after the first P arrival for TEAM.
We use a longer time before the first prediction for EPS as the early results of EPS are unstable.
Warnings are not retracted, i.e., even if the model later estimates a shake level below the warning threshold, the warning stay active.

\clearpage

\renewcommand\thefigure{S\arabic{figure}}    
\setcounter{figure}{0}    

\renewcommand\thetable{S\arabic{table}}    
\setcounter{table}{0}  

\begin{figure}
 \includegraphics[width=\textwidth]{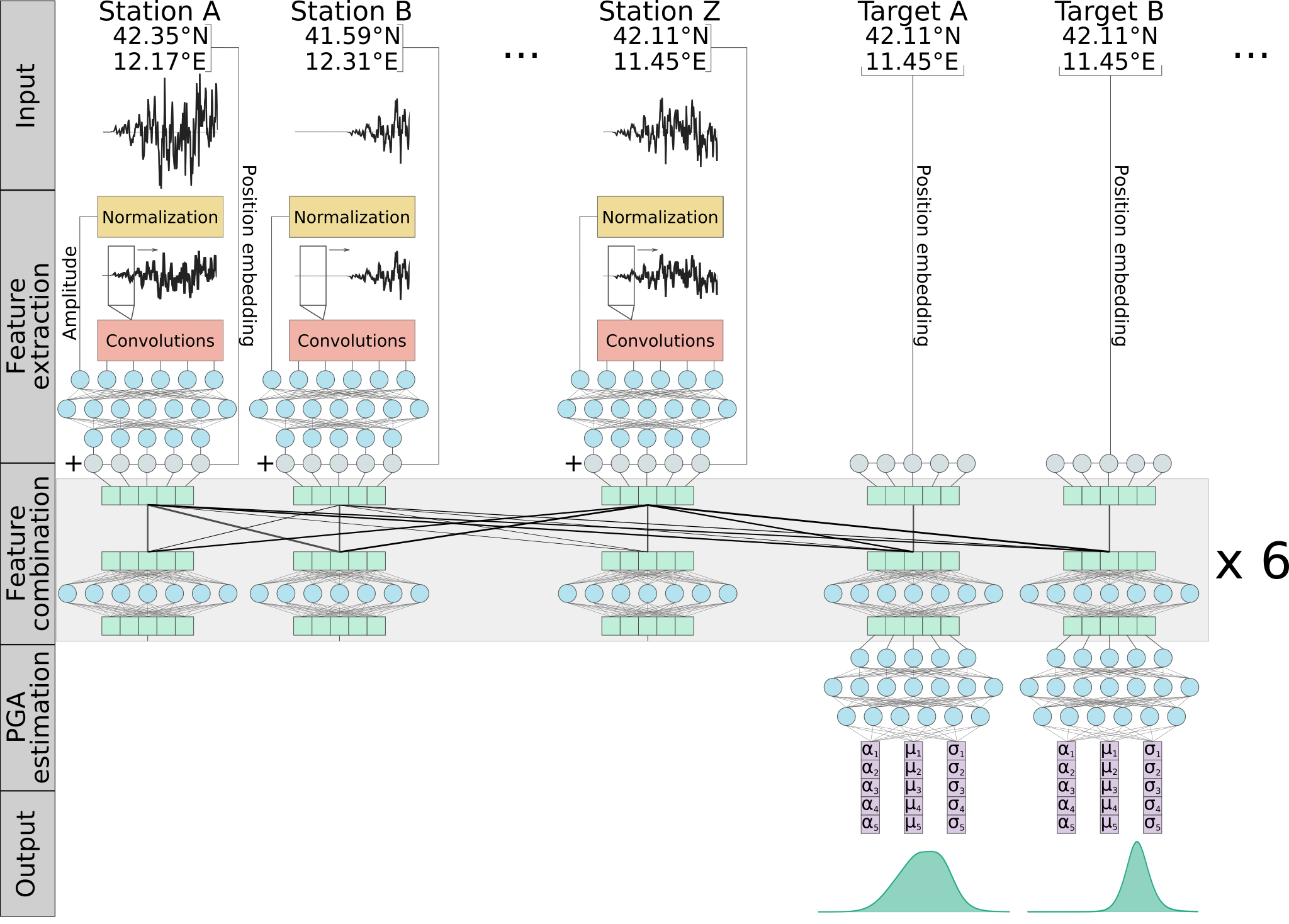}
 \caption{Overview of the transformer earthquake alerting model, showing the input, the feature extraction, the feature combination, the PGA estimation and the output. For simplicity, not all layers are shown, but only their order and combination is visualized schematically. For the exact number of layers and the size of each layer please refer to tables \ref{tab:convolutions} and \ref{tab:transformer}. Please note that the number of input stations and the number of targets are both variable, due to the self-attention mechanism in the feature combination. Ten instances of this network are trained independently and the results ensemble-averaged.}
 \label{fig:model_overview}
\end{figure}

\begin{figure}
 \centering
 \includegraphics[width=0.83\textwidth]{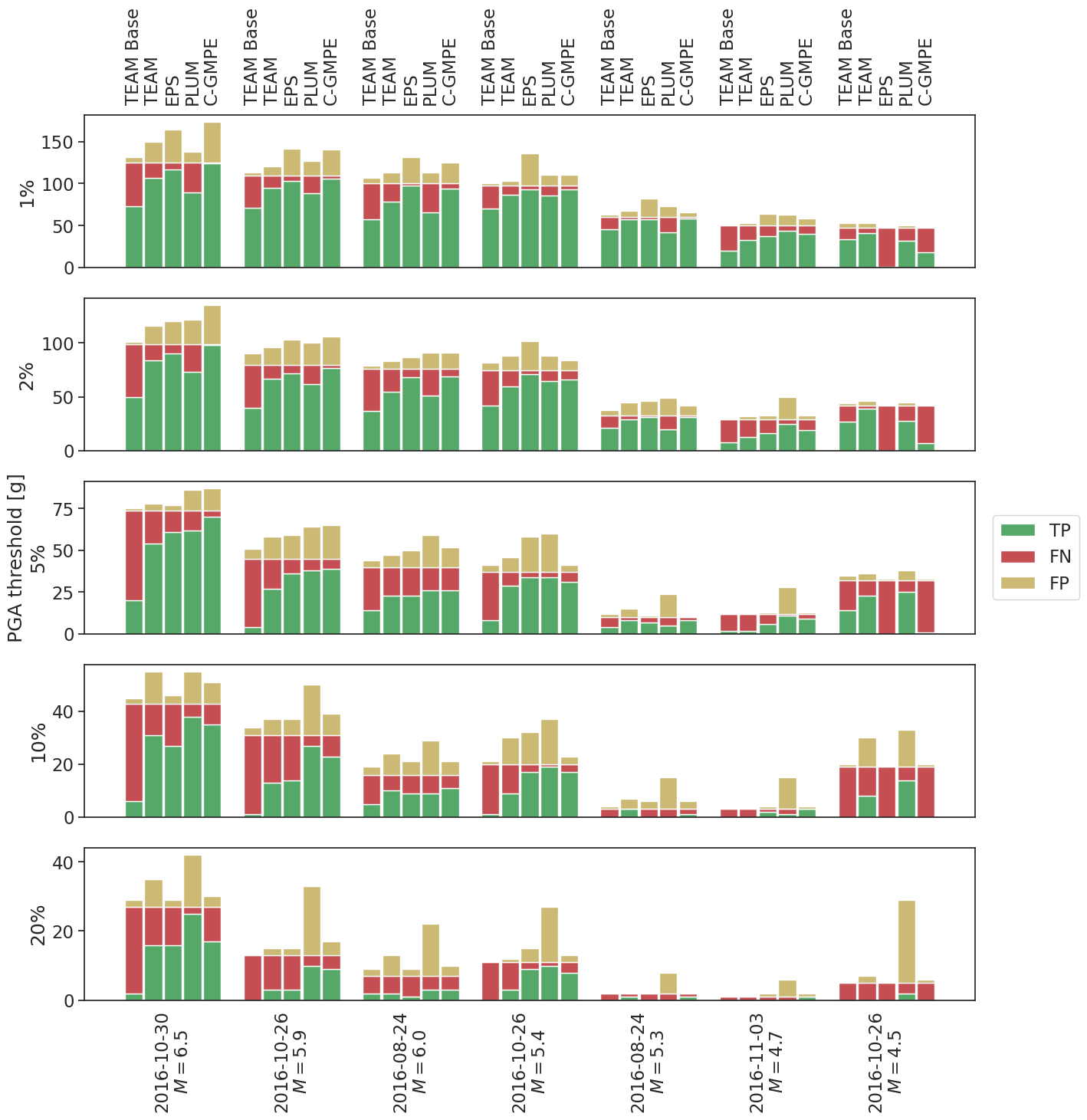}
 \caption{True positives (TP), false negatives (FN) and false positives (FP) for the events in the Italy test sets causing the largest shaking. The methods are the transformer earthquake alerting model without domain adaptation (TEAM base), the transformer earthquake alerting model (TEAM), the estimated point source algorithm (EPS) and PLUM-based approach. In addition, a GMPE with full catalog information is included for reference. Values $\alpha$ were chosen separately for each threshold and method to yield the  highest F1 score for the whole test set, but are kept constant across all events. TEAM with domain adaptation outperforms TEAM without domain adaptation consistently across all thresholds. This indicates that the domain adaptation not only allows TEAM to better predict higher levels of shaking, but also to better assess large events in general.}
 \label{fig:tpfpfn_large_events}
\end{figure}

\begin{figure}
 \includegraphics[width=\textwidth]{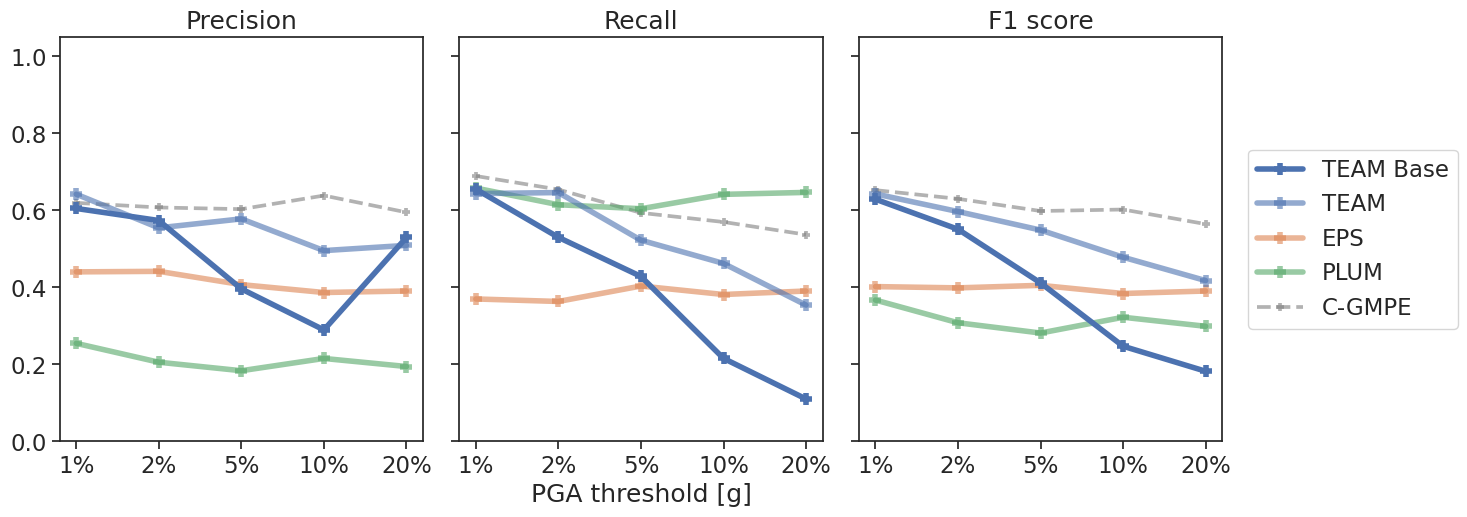}
 \caption{Precision, recall and F1 score at different PGA thresholds for Italy including TEAM without domain adaptation. Threshold values $\alpha$ were chosen independently for each method and PGA threshold to yield the highest F1 score. The methods are the transformer earthquake alerting model without domain adaptation (TEAM Base), the transformer earthquake alerting model (TEAM), the estimated point source (EPS) model and the PLUM-based model. In addition the graph shows the performance of C-GMPE, a GMPE with full catalog information for reference.}
 \label{fig:results_thresholds_plain}
\end{figure}

\begin{figure}
 \includegraphics[width=\textwidth]{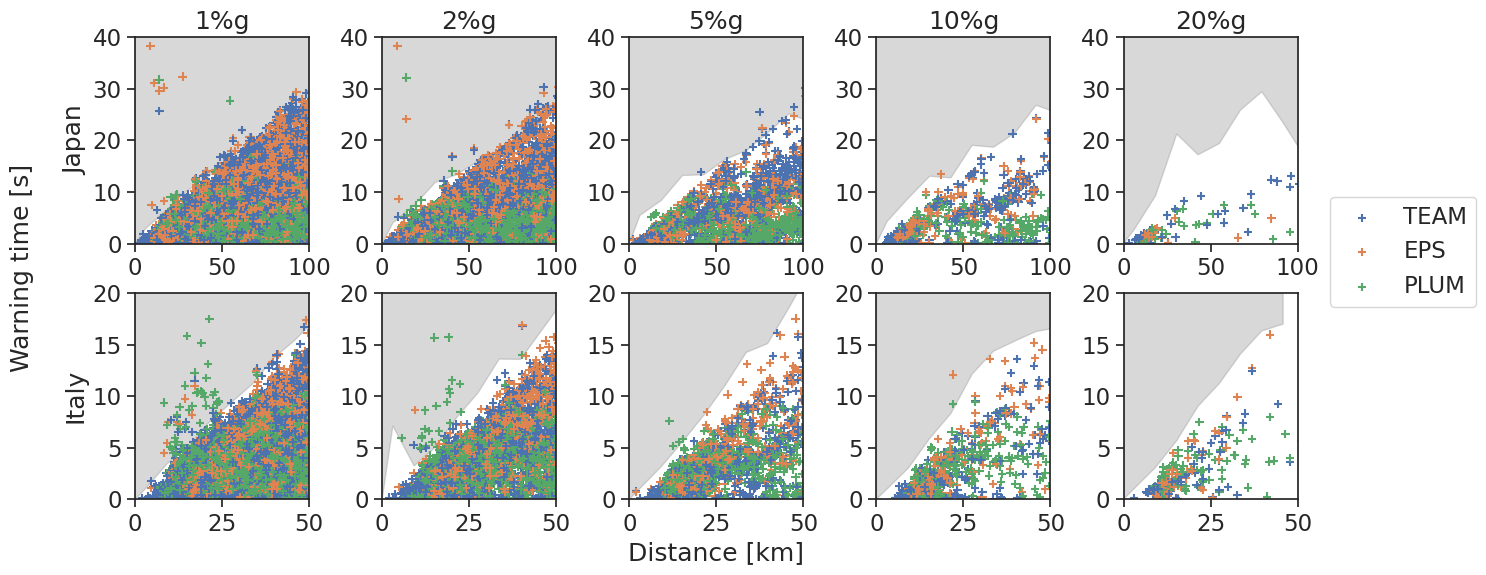} 
 \caption{Warning time and hypocentral distance between station and event for each true alert at F1-optimal $\alpha$. The white area corresponds roughly to the range of possible warning times and is bounded by the 90$^{th}$ percentile of the times between first detection of an event (i.e., arrival of P wave at the closest station) and first exceedance of the PGA threshold in recordings at that approximate distance.}
 \label{fig:warning_dist}
\end{figure}

\begin{figure}
 \includegraphics[width=\textwidth]{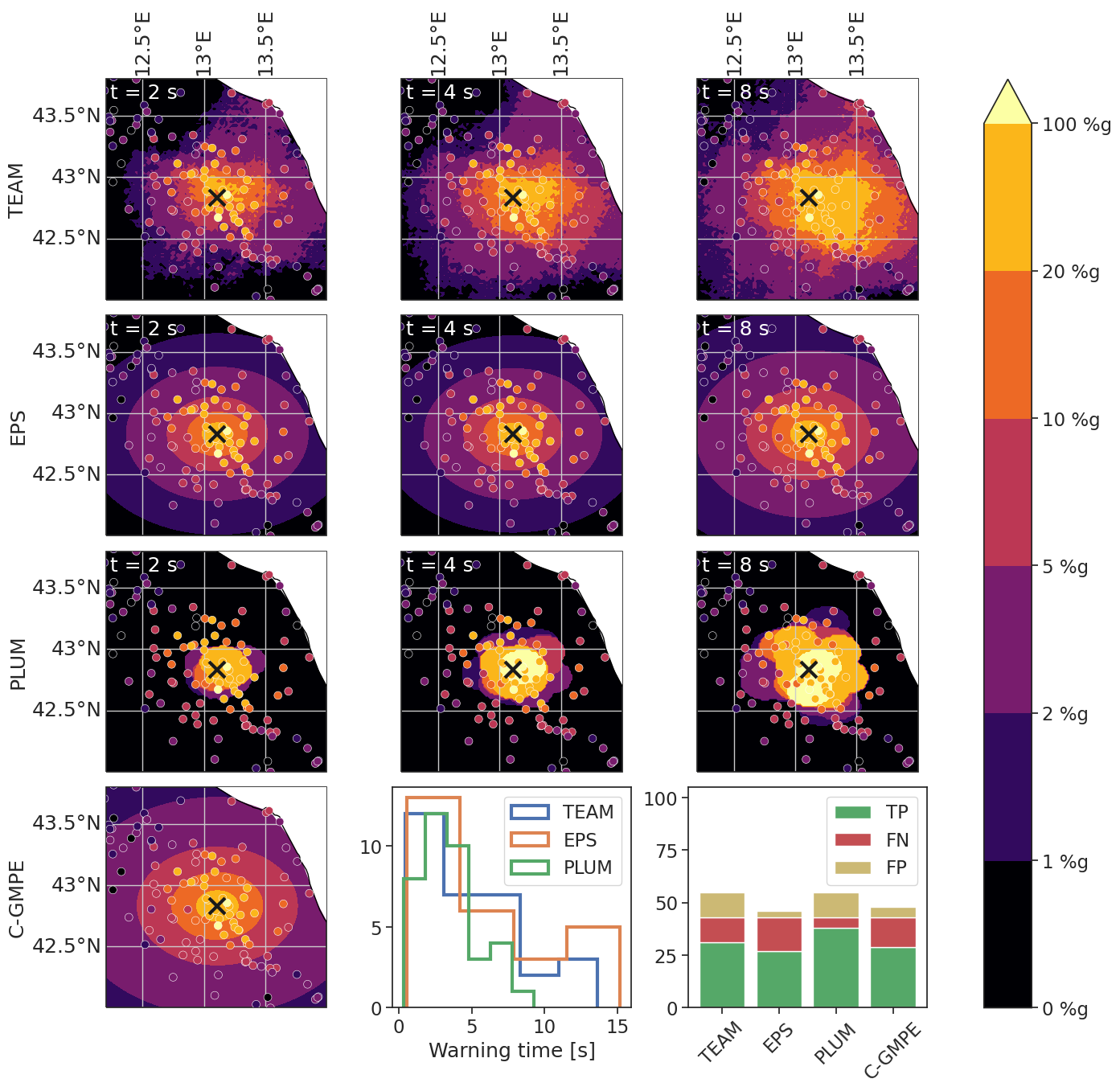}
 \caption{Scenario analysis of the 30th October 2016 $M_w=6.5$ Norcia earthquake, the largest event in the Italy test set.  See Fig.~4 in the main paper for further explanations. The bottom row diagrams for this scenario analysis use a 10\%g PGA threshold.}
 \label{fig:scenario_norcia}
\end{figure}

\begin{figure}
 \includegraphics[width=\textwidth]{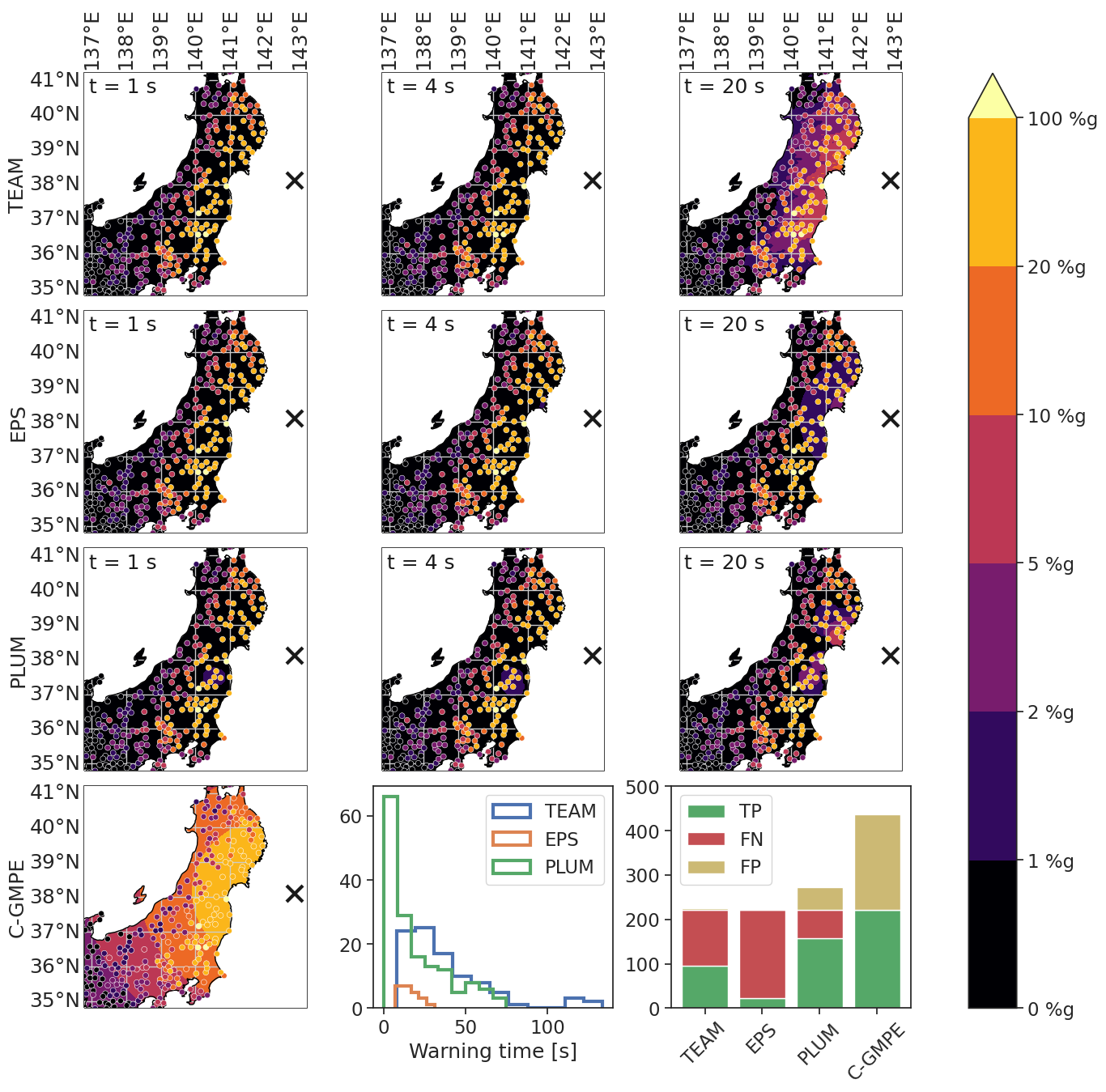}
 \caption{Scenario analysis of the 11th March 2011 $M_w=9.1$ Tohoku earthquake, the largest event in the Japan dataset. See Fig.~4 in the main paper for further explanations. The bottom row diagrams for this scenario analysis use a 2\%g PGA threshold.}
 \label{fig:scenario_tohoku}
\end{figure}

\begin{figure}
 \centering
 \includegraphics[width=\textwidth]{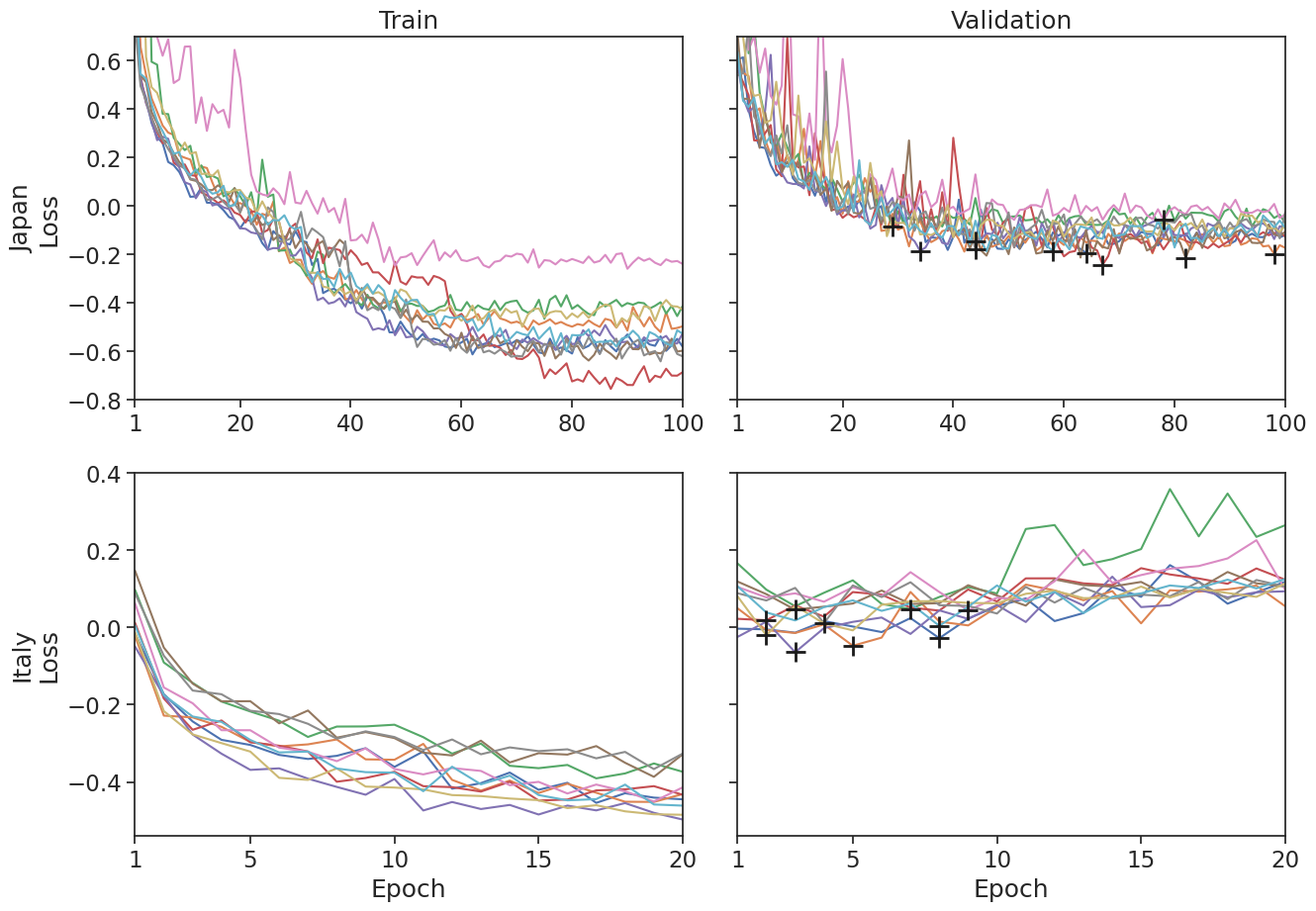}
 \caption{Training and validation loss curves for the Japan TEAM model and the fine-tuning step of the Italy TEAM model. Each line shows the loss curve for one ensemble member with colors matching between training and validation curves. The models used are determined by the minimum validation loss and are denoted by black crosses. The models were evaluated after the training epoch indicated on the x-axis, i.e., the leftmost point of each curve already includes one epoch of training.}
 \label{fig:loss_curves}
\end{figure}

\begin{figure}
 \centering
 \includegraphics[width=0.9\textwidth]{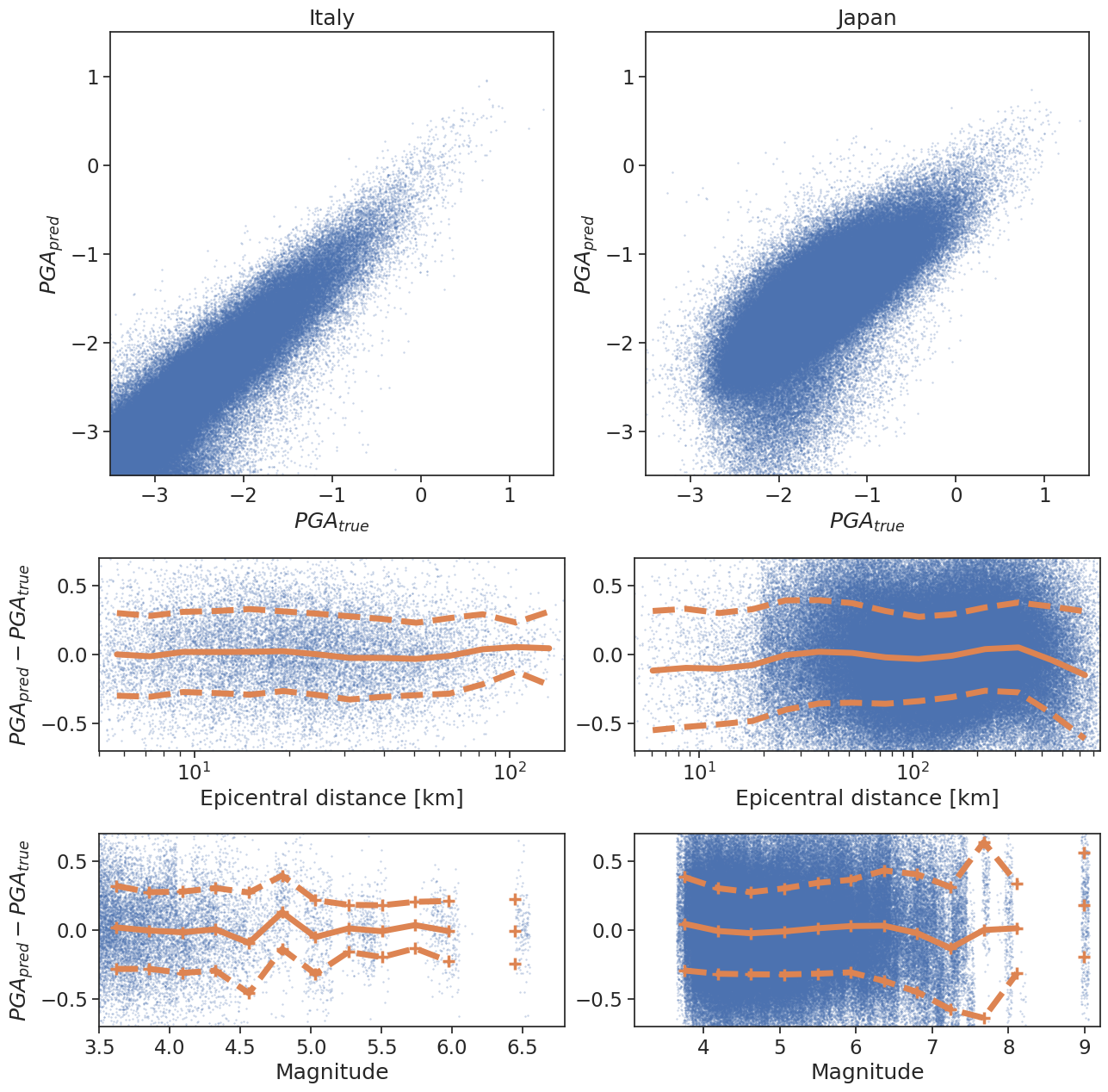}
 \caption{Predictions and residuals of the GMPEs derived in this study. All PGA values are given as log units using $m/s^2$. Every point refers to one recording. Solid lines indicate running means, dashed lines denote the running standard deviation around the running mean. Orange crosses denote mean and standard deviations for magnitude ranges with insufficient data to infer a continuous line. Window sizes are 0.24~m.u./10~km (Italy) and 0.44~m.u./53~km (Japan). Overall $\sigma$ is 0.29 for Italy and 0.33 for Japan. The plotted magnitude values have been offset by random values between -0.05 and 0.05 m.u. for increased visibility.}
 \label{fig:residuals}
\end{figure}

\begin{figure}
 \includegraphics[width=\textwidth]{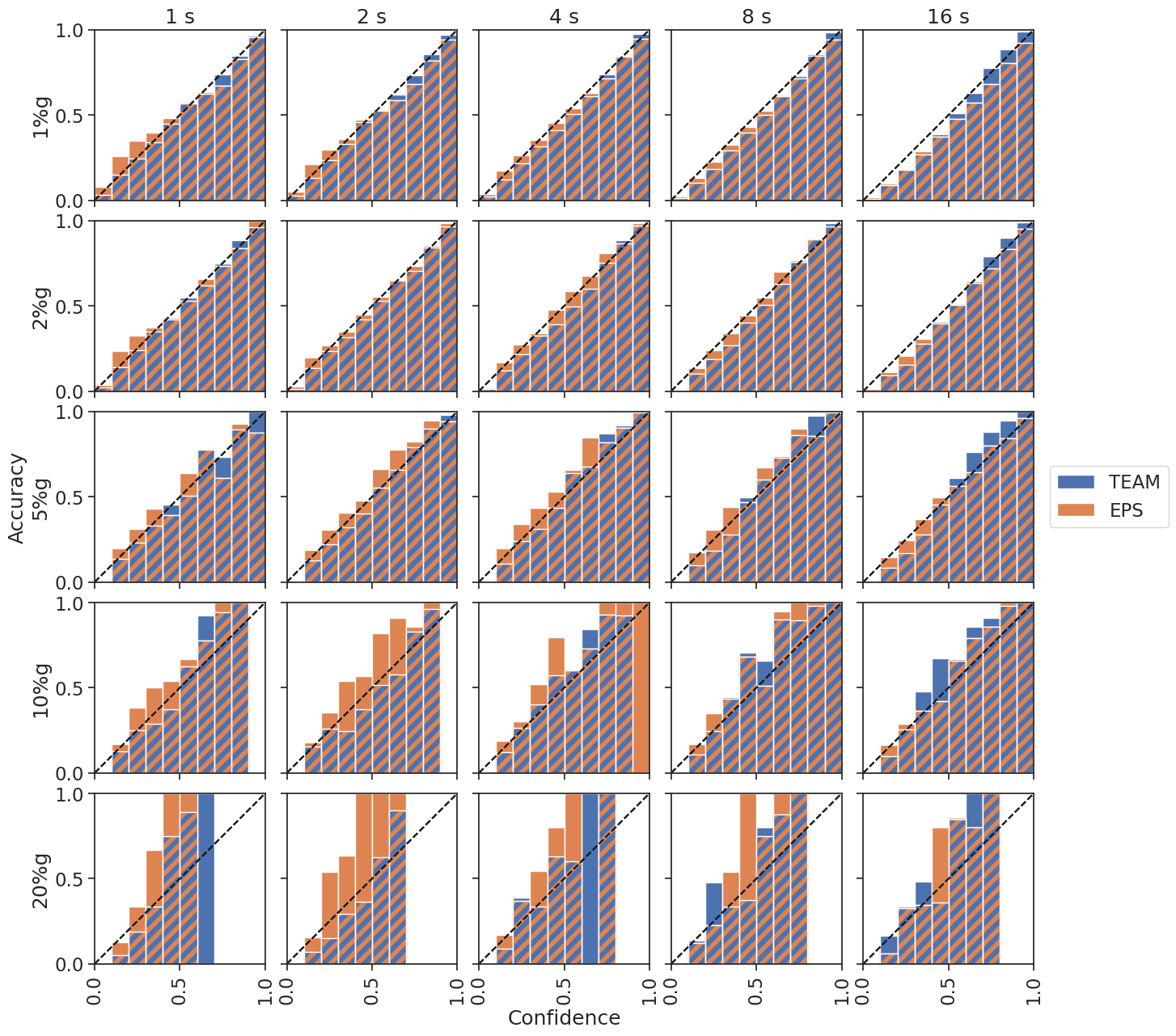}
 \caption{Calibration diagrams for Japan at different times after the first P detection and different PGA thresholds. The confidence is defined as the probability of exceeding the PGA threshold as predicted by the model. Each bar represents the traces with a confidence value inside the limits of the bar. Its height is given by the accuracy, the fraction of traces actually exceeding the threshold among all traces in the bar. For a perfectly calibrated model, the confidence equals the accuracy. This is indicated by the dashed line. We note that accuracy estimations for the high PGA thresholds are strongly impacted by stochasticity due to the small number of samples.}
 \label{fig:calibration_japan}
\end{figure}

\begin{figure}
 \includegraphics[width=\textwidth]{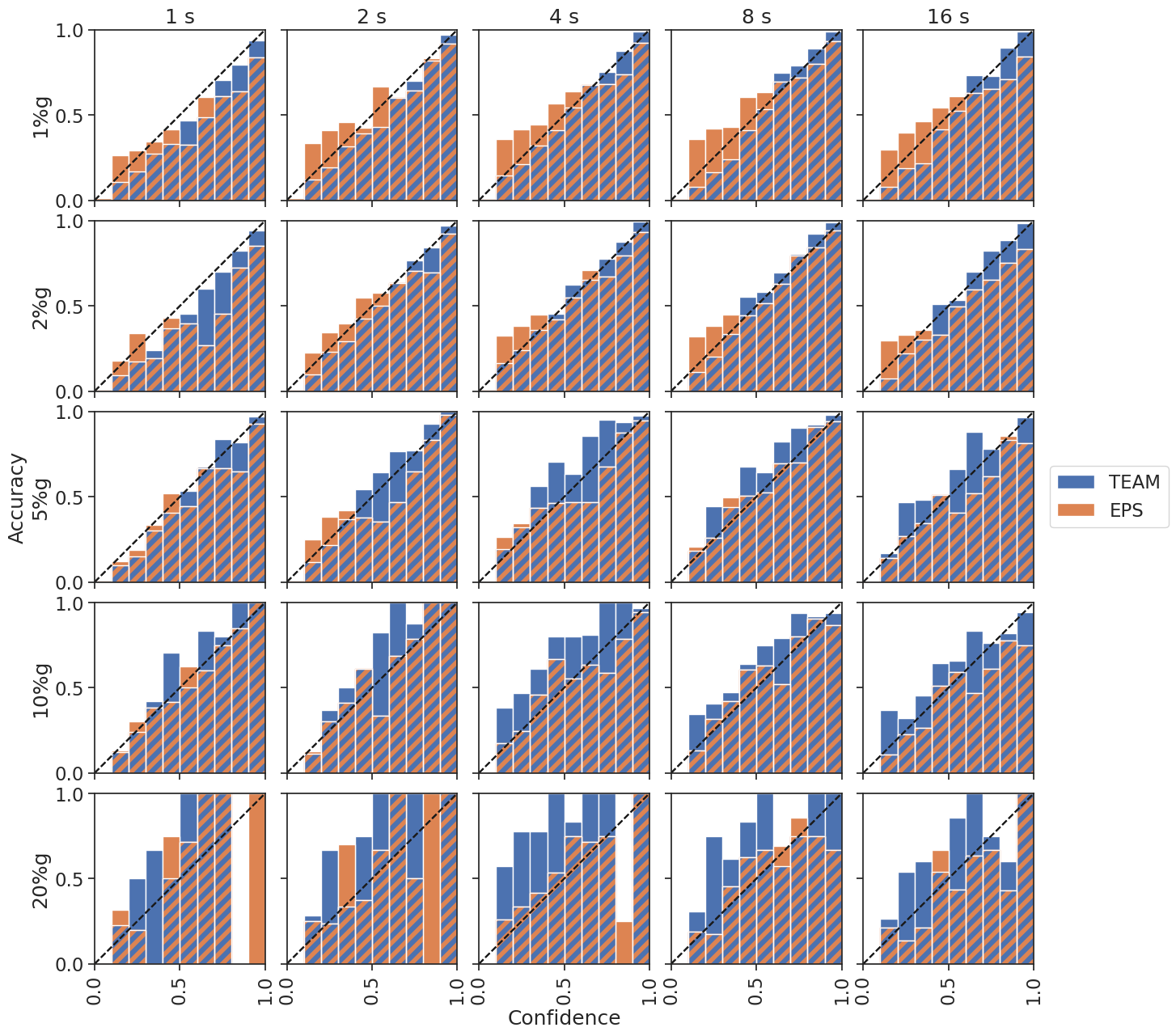}
 \caption{Calibration diagrams for Italy at different times after the first P detection and different PGA thresholds. For a further description see the caption of figure \ref{fig:calibration_japan}.}
 \label{fig:calibration_italy}
\end{figure}

\begin{table}
 \centering
 \caption{Performance statistics for Japan. Probability thresholds $\alpha$ were chosen to maximize F1 scores and are shown in the last column. The AUC value does not depend on the threshold $\alpha$. PGA indicates the used PGA threshold.}
 \input{performance_stats_japan.tex}
 \label{tab:performance_stats_japan}
\end{table}

\begin{table}
 \centering
 \caption{Performance statistics for Italy. Probability thresholds $\alpha$ were chosen to maximize F1 scores and are shown in the last column. The AUC value does not depend on the threshold $\alpha$. PGA indicates the used PGA threshold.}
 \input{performance_stats_italy.tex}
 \label{tab:performance_stats_italy}
\end{table}

\begin{table}
 \centering
 \caption{Relative warning times of the algorithms in seconds. Positive values indicate longer average warning times for the second method, negative values shorter warning times. The difference in average warning times is calculated from all event station pairs, where both methods issued correct warnings. No value is reported if this set is empty. We set $\alpha$ for TEAM and EPS to the optimal value in terms of F1 score.}
 \input{warning_times.tex}
 \label{tab:warning_times}
\end{table}

\begin{table}
 \centering
 \caption{Data set statistics for the full data set and the test set. The lower boundary of the magnitude category is the 5th percentile of the magnitude; this limit is chosen as each data set contains a small number of unrepresentative very small events. The upper boundary is the maximum magnitude. The lower part of the table shows how often each PGA threshold was exceeded. An event is counted as exceeding a threshold if at least one station exceeded this threshold during the event. The number of exceedances in the test set for Italy is disproportionally high compared to the number of events in the test set. This is caused by the high seismic activity and the higher station density in 2016. Traces for Japan always refer to 6 component traces, while for Italy it refers to 3 component traces.}
 \input{datasets.tex}
 \label{tab:datasets}
\end{table}

\begin{table}
 \centering
 \caption{Architecture of the feature extraction network. The input dimensions of the waveform data are (time, channels). FC denotes fully connected layers. As FC layers can be regarded as 0D convolutions, we write the output dimensionality in the filters column. The ``Concatenate scale'' layer concatenates the log of the peak amplitude to the output of the convolutions. Depending on the existence of borehole data the number of input filters for the first Conv1D layer is 64 instead of 32 in the non-borehole case.}
 \input{convolutions.tex}
 \label{tab:convolutions}
\end{table}

\begin{table}
 \centering
 \caption{Architecture of the transformer network. Please note that even though the transformer in TEAM does not apply dropout, we explicitly state this in the table, as transformers commonly use dropout.}
 \input{transformer.tex}
 \label{tab:transformer}
\end{table}

\end{document}

%% file: performance_stats_japan.tex
\begin{tabular}{ccccccc}
& PGA [g] & Precision & Recall & F1 & AUC & $\alpha$ \\\hline
\multirow{5}{*}{TEAM}
& 1\% & 0.70 & 0.77 & 0.73 & 0.82 & 0.60 \\
& 2\% & 0.69 & 0.69 & 0.69 & 0.76 & 0.60 \\
& 5\% & 0.59 & 0.67 & 0.63 & 0.68 & 0.50 \\
& 10\% & 0.50 & 0.60 & 0.54 & 0.56 & 0.40 \\
& 20\% & 0.33 & 0.48 & 0.39 & 0.35 & 0.30 \\
\hline
\multirow{5}{*}{EPS}
& 1\% & 0.50 & 0.63 & 0.56 & 0.57 & 0.40 \\
& 2\% & 0.48 & 0.48 & 0.48 & 0.48 & 0.40 \\
& 5\% & 0.40 & 0.40 & 0.40 & 0.34 & 0.30 \\
& 10\% & 0.27 & 0.36 & 0.31 & 0.25 & 0.20 \\
& 20\% & 0.20 & 0.26 & 0.22 & 0.15 & 0.20 \\
\hline
\multirow{5}{*}{PLUM}
& 1\% & 0.39 & 0.56 & 0.46 & - & - \\
& 2\% & 0.30 & 0.50 & 0.38 & - & - \\
& 5\% & 0.22 & 0.42 & 0.29 & - & - \\
& 10\% & 0.18 & 0.39 & 0.25 & - & - \\
& 20\% & 0.11 & 0.28 & 0.16 & - & - \\
\hline
\multirow{5}{*}{C-GMPE}
& 1\% & 0.58 & 0.74 & 0.65 & 0.69 & 0.30 \\
& 2\% & 0.47 & 0.71 & 0.56 & 0.60 & 0.20 \\
& 5\% & 0.44 & 0.54 & 0.48 & 0.48 & 0.20 \\
& 10\% & 0.44 & 0.46 & 0.45 & 0.43 & 0.20 \\
& 20\% & 0.56 & 0.38 & 0.45 & 0.42 & 0.30 \\
\end{tabular}

%% file: performance_stats_italy.tex
\begin{tabular}{ccccccc}
& PGA [g] & Precision & Recall & F1 & AUC & $\alpha$ \\\hline
\multirow{5}{*}{TEAM}
& 1\% & 0.64 & 0.64 & 0.64 & 0.68 & 0.60 \\
& 2\% & 0.55 & 0.65 & 0.60 & 0.63 & 0.50 \\
& 5\% & 0.58 & 0.52 & 0.55 & 0.54 & 0.50 \\
& 10\% & 0.50 & 0.46 & 0.48 & 0.43 & 0.40 \\
& 20\% & 0.51 & 0.35 & 0.42 & 0.36 & 0.30 \\
\hline
\multirow{5}{*}{EPS}
& 1\% & 0.44 & 0.37 & 0.40 & 0.37 & 0.30 \\
& 2\% & 0.44 & 0.36 & 0.40 & 0.36 & 0.40 \\
& 5\% & 0.41 & 0.40 & 0.40 & 0.33 & 0.40 \\
& 10\% & 0.39 & 0.38 & 0.38 & 0.30 & 0.40 \\
& 20\% & 0.39 & 0.39 & 0.39 & 0.25 & 0.40 \\
\hline
\multirow{5}{*}{PLUM}
& 1\% & 0.25 & 0.66 & 0.37 & - & - \\
& 2\% & 0.21 & 0.61 & 0.31 & - & - \\
& 5\% & 0.18 & 0.60 & 0.28 & - & - \\
& 10\% & 0.22 & 0.64 & 0.32 & - & - \\
& 20\% & 0.19 & 0.65 & 0.30 & - & - \\
\hline
\multirow{5}{*}{C-GMPE}
& 1\% & 0.62 & 0.69 & 0.65 & 0.71 & 0.30 \\
& 2\% & 0.61 & 0.65 & 0.63 & 0.68 & 0.30 \\
& 5\% & 0.60 & 0.59 & 0.60 & 0.63 & 0.30 \\
& 10\% & 0.64 & 0.57 & 0.60 & 0.59 & 0.30 \\
& 20\% & 0.59 & 0.54 & 0.56 & 0.54 & 0.30 \\
\end{tabular}

%% file: warning_times.tex
\begin{tabular}{cc|ccccc|ccccc}
& & \multicolumn{5}{c|}{Japan} & \multicolumn{5}{c}{Italy} \\
\multicolumn{2}{c|}{PGA [g]} & 1\% & 2\% & 5\% & 10\% & 20\% & 1\% & 2\% & 5\% & 10\% & 20\% \\
\hline
EPS & TEAM & 0.39 & 0.43 & 0.70 & 0.31 & 0.61 & 0.18 & 0.26 & -0.49 & -0.65 & -1.19 \\
PLUM & TEAM & 8.98 & 8.24 & 6.35 & 5.01 & 0.55 & 1.49 & 1.60 & 1.03 & -0.03 & 0.03 \\
PLUM & EPS & 8.53 & 7.74 & 5.29 & 3.08 & -0.04 & 2.95 & 3.11 & 2.35 & 0.81 & 1.08 \\
\end{tabular}

%% file: datasets.tex
\begin{tabular}{c|cccc|cccc}
& \multicolumn{4}{c|}{Japan} & \multicolumn{4}{c}{Italy} \\
& \multicolumn{2}{c}{Full} & \multicolumn{2}{c|}{Test} & \multicolumn{2}{c}{Full} & \multicolumn{2}{c}{Test} \\
\hline
Years & \multicolumn{2}{c}{1997 - 2018} & \multicolumn{2}{c|}{08/2013 - 12/2018} & \multicolumn{2}{c}{2008 - 2019} & \multicolumn{2}{c}{01/2016 - 12/2016} \\
Magnitudes & \multicolumn{2}{c}{2.7 - 9.0} & \multicolumn{2}{c|}{2.7 - 8.1} & \multicolumn{2}{c}{2.7 - 6.5} & \multicolumn{2}{c}{2.7 - 6.5} \\
Events & \multicolumn{2}{c}{13,512} & \multicolumn{2}{c|}{4,054} & \multicolumn{2}{c}{7,055} & \multicolumn{2}{c}{2,123} \\
Unique stations & \multicolumn{2}{c}{697} & \multicolumn{2}{c|}{632} & \multicolumn{2}{c}{1,080} & \multicolumn{2}{c}{621} \\
Traces & \multicolumn{2}{c}{372,661} & \multicolumn{2}{c|}{104,573} & \multicolumn{2}{c}{494,183} & \multicolumn{2}{c}{253,454} \\
Avg. traces per event & \multicolumn{2}{c}{27.6} & \multicolumn{2}{c|}{25.9} & \multicolumn{2}{c}{70.3} & \multicolumn{2}{c}{119.4} \\
\hline
PGA [g] & Events & Traces & Events & Traces & Events & Traces & Events & Traces \\
1\% & 8,761 & 55,618 & 2,710 & 15,215 & 1,841 & 6,379 & 923 & 3,826 \\
2\% & 5,324 & 24,396 & 1,601 & 6,489 & 1,013 & 2,921 & 503 & 1,771 \\
5\% & 2,026 & 6,802 & 583 & 1,712 & 348 & 888 & 171 & 563 \\
10\% & 782 & 2,223 & 216 & 506 & 120 & 330 & 58 & 223 \\
20\% & 238 & 631 & 62 & 100 & 40 & 107 & 20 & 82 \\
\end{tabular}

%% file: convolutions.tex
\begin{tabular}{cccc}
Layer & Filters & Kernel size & Stride \\
\hline
Conv2D & 8 & 5, 1 & 5, 1 \\
Conv2D & 32 & 16, 3 & 1, 3 \\
Flatten to 1D & & & \\
Conv1D & 64 & 16 & 1 \\
MaxPool1D & & 2 & 2 \\
Conv1D & 128 & 16 & 1 \\
MaxPool1D & & 2 & 2 \\
Conv1D & 32 & 8 & 1 \\
MaxPool1D & & 2 & 2 \\
Conv1D & 32 & 8 & 1 \\
Conv1D & 16 & 4 & 1 \\
Flatten to 0D& & & \\
Concatenate scale & & & \\
FC & 500 & & \\
FC & 500 & & \\
FC & 500 & & \\
\end{tabular}

%% file: transformer.tex
\begin{tabular}{cc}
Feature & Value \\
\hline
\# Layers & 6 \\
Dimension & 500 \\
Feed forward dimension & 1000 \\
\# Heads & 10 \\
Maximum number of stations & 25 \\
Dropout & 0 \\
Activation & GeLu \\
\end{tabular}